\documentclass[desactivate]{aa} 

\usepackage{ulem}
\usepackage{graphicx}
\usepackage{siunitx}
\usepackage{booktabs}
\usepackage{multirow}
\usepackage{subcaption}
\usepackage{tikz}
\usetikzlibrary{shapes.geometric, arrows.meta, positioning}
\usepackage[varg]{txfonts}
\usepackage{xcolor}
\definecolor{darkblue}{rgb}{0.0, 0.0, 0.55}
\definecolor{darkgreen}{rgb}{0.0, 0.4, 0.0}
\usepackage[pagebackref=false,colorlinks=true,urlcolor=blue,citecolor=blue,linkcolor=blue,hyperindex=true]{hyperref}
\usepackage{makecell}
\usepackage{orcidlink}
\usepackage{placeins}

\DeclareMathOperator{\doop}{\mathrm{do}}
\DeclareMathOperator{\shift}{\mathrm{shift}}

\DeclareSIUnit\angstrom{\text {Å}}

\begin{document}
   \title{Stellar parameter prediction and spectral simulation using machine learning \thanks{Based on data obtained from the ESO Science Archive Facility with DOI(s): \url{https://doi.eso.org/10.18727/archive/33}}}

   \subtitle{A systematic comparison of methods with HARPS observational data}

   \author{Vojt\v{e}ch Cvr\v{c}ek
          \inst{1, 2}{\orcidlink{0000-0002-2552-6952}}
          \and
          Martino Romaniello
          \inst{2}\orcidlink{0000-0002-5527-6317}
          \and
          Radim \v{S}\'{a}ra
          \inst{1}\orcidlink{0000-0002-2032-5764}
          \and
          Wolfram Freudling
          \inst{2}\orcidlink{0000-0002-7941-9554}
          \and
          Pascal Ballester 
          \inst{2}\orcidlink{0009-0000-4734-6154}
          }
    \authorrunning{Vojt\v{e}ch Cvr\v{c}ek et al.}

   \institute{
                Department of Cybernetics, Czech Technical University in Prague, Czech Republic\\
                \email{cvrcevo1@fel.cvut.cz}
                \and
                European Southern Observatory,
                Karl-Schwarzschild-Str. 2,
                85748 Garching,
                Germany
            }

     \date{Received date June 16, 2024 /
       Accepted date November 4, 2024}

 
  \abstract
   {}
    {We applied machine learning to the entire data history of ESO’s High Accuracy Radial Velocity Planet Searcher (HARPS) instrument.
    Our primary goal was to recover the physical properties of the observed objects, with a secondary emphasis on simulating spectra.
    We systematically investigated the impact of various factors on the accuracy and fidelity of the results, including the use of simulated data, the effect of varying amounts of real training data, network architectures, and learning paradigms.
}
    {Our approach integrates supervised and unsupervised learning techniques within autoencoder frameworks.
    Our methodology leverages an existing simulation model that utilizes a library of existing stellar spectra in which the emerging flux is computed from first principles rooted in physics and a HARPS instrument model to generate simulated spectra comparable to observational data.
    We trained standard and variational autoencoders on HARPS data to predict spectral parameters and generate spectra.
    Convolutional and residual architectures were compared, and we decomposed autoencoders in order to assess component impacts.
    }
   {Our models excel at predicting spectral parameters and compressing real spectra, and they achieved a mean prediction error of $\sim50$~K for effective temperatures, making them relevant for most astrophysical applications.
    Furthermore, the models predict metallicity ([M/H]) and surface gravity ($\log g$) with an accuracy of $\sim0.03$~dex and $\sim0.04$~dex, respectively, underscoring their broad applicability in astrophysical research.
    Moreover, the models can generate new spectra that closely mimic actual observations, enriching traditional simulation techniques.
    Our variational autoencoder-based models achieve short processing times: 779.6 \si{\milli\second} on a CPU and 3.97 \si{\milli\second} on a GPU.
    These results demonstrate the benefits of integrating high-quality data with advanced model architectures, as it significantly enhances the scope and accuracy of spectroscopic analysis. With an accuracy comparable to the best classical analysis method but requiring a fraction of the computation time, our methods are particularly suitable for high-throughput observations such as massive spectroscopic surveys and large archival studies.
    }
   {}

   \maketitle
%

\section{Introduction}
\label{sec:intro}
Deriving reliable source parameters from very large sets of spectra has been rapidly gaining importance as the amount of such data massively increases, for example, through dedicated observational campaigns and/or data piling up in public archives. Perhaps the most notable case is the analysis of spectra from the ESA \textit{Gaia} mission, as it has provided the largest homogeneously observed stellar spectral sample to date \citep[470 million stars with spectra in Data Release 3;][]{gaiacollaborationGaiaDataRelease2022}.

Manual analyses of such large datasets are ``de facto'' impossible, and automatic techniques need to take their place.
As with any data analysis, it is imperative that the associated uncertainties, random and systematic, as well as the limitations of the methodologies are thoroughly understood and quantified so that the results can be reliably used, including by researchers who have not participated in their generation.
Also, the methods need to be computationally efficient in order to cope with the correspondingly large data volumes in terms of the number of individual spectra and their sheer size.
This, of course, is in addition to the need to deliver precision and accuracy that are competitive with the best results available in the literature.

In this research, we use machine learning (ML) to interpret spectral data, focusing on key physical and observational parameters as our primary interest.
These spectral parameters include temperature, surface gravity, radial velocity, metallicity, airmass, and barycentric Earth radial velocity.
Within our ML framework, these spectral parameters are treated as ``labels,'' and they are the outcomes our model is designed to predict.

We chose the public catalog of observations from the HARPS instrument at the ESO La Silla Observatory for our study.
This dataset is accessible through the ESO Science Archive\footnote{\url{https://archive.eso.org}} and is continually updated as the instrument remains in active operation.
As per the nature of the instrument, the targets are stars in the solar neighborhood mostly originally observed with the intention of searching for planets around them.
Our study does not focus on this particular aspect but instead on the determination of the physical parameters of the stars.
The sample is particularly well suited for this since given the relatively small distance from Earth to the targets, they have been well studied in the literature, and thus there is an extensive set of physical parameters that can be used as labels to train the networks with.
Also, the spectra themselves are extremely rich in information. Due to the high spectral resolution and broad wavelength coverage of HARPS, a normal stellar spectrum displays hundreds of features (absorption lines).
 
Machine learning techniques have experienced a rise in popularity in the past several years, being promising tools for analyzing astronomical spectra (as well as other types of astronomical data).
Once trained, the methods can be applied to large datasets, usually in a computationally efficient way.
Many trained models exist, but their inner workings are not always clear, so they frequently resemble black boxes.
Moreover, distinct techniques are frequently utilized for specific datasets, which complicates the disentanglement of their individual roles and impacts.

Typically, ML algorithms offer a large set of customizable options that must be tuned to get optimal results. This can be a confusing and daunting task.
Here, we have pursued a systematic approach by comparing different ML algorithms and setups and applying them to the same dataset.
Our goal is to find a set of principles regarding how to navigate through these options and propose a practical methodology that can be applied to other similar investigations. To achieve this goal, we aim to provide objective and reproducible results, tracing the effects of the input assumptions and setup choices on the output by exploring a representative set of techniques and ML parameters.
Crucially, these techniques and parameters include the number of labels available for training. Since labels may not always be copiously accessible, they will be, in many cases, the limiting factor to the attainable accuracy.

Our primary objective is to predict stellar parameters, a task we call ``label prediction.''
Our secondary objective is to generate realistic synthetic spectra, a task we call ``ML simulation,'' to clearly distinguish from the traditional physics-based simulations based on solving the transfer equations in the stellar atmospheres.
We are interested in solving these tasks jointly with models that can harness multiple sources of information: the data itself, in order to circumvent the challenging task of obtaining enough reliable labels for supervised learning; catalog data, to form a semantically meaningful latent space directly; and synthetic data with uniform distributions of spectral parameters, which can serve as an alternative strategy to deal with a lack of labels.

We work under the assumption that label prediction and ML simulation are related tasks.
Therefore, a joint model capable of performing both tasks simultaneously could outperform models that specialize in these tasks separately.
Lastly, we hypothesize that the model can benefit from the additional information provided by the synthetic data, which can be used to regularize the model and improve its generalization capabilities.

The paper is organized as follows. Section~\ref{sec:methods} presents the problem and explains in detail the methods used. Section~\ref{sec:experiments} covers the data, its augmentation, our learning approach, and the metrics employed.
Section~\ref{sec:results} presents the experimental results, while Sect.~\ref{sec:discussion} offers a discussion of these results. Finally, Sect.~\ref{sec:conclusions} summarizes the conclusions of the paper and explores possible future research directions.

\section{Machine learning methods}
\label{sec:methods}
This section provides a concise overview of the ML models and methods we employ to achieve our objectives, as well as alternative, state-of-the-art models.
We begin by defining fundamental ML terminology and mathematical notation relevant to our experiments.
Next, we clarify our objectives by defining the inference tasks for label prediction and ML simulation.
Subsequently, we formalize the learning problem associated with these inference tasks.
Finally, we introduce specific ML models that we use to address the inference tasks.

\subsection{Machine learning preliminaries}
In this section we explain the relevant concepts and terminology that are essential to understand the following text.
We also investigate the application of ML techniques for label prediction and ML simulations.
We specifically examine the use of interpretable unsupervised models and how to combine them with supervision to enhance performance. 

We start with classic ``supervised learning'' \citep[p.~1]{Murphy_22} as a basic framework to extract labels from spectral data.
This involves the application of an ML model learned from data annotated by experts, which can then be used to predict parameters for previously unseen data.
Supervised learning has seen wide-ranging applications in astronomy, such as galaxy morphology classification \citep{morpho_class}, star-galaxy separation \citep{star_galaxy}, and transient detection and classification \citep{light_curves}. These works often employ methods such as random forests and neural networks.
An overview of supervised learning in astronomy is presented in \citep{astro_ML_overview}.

The estimation of physical parameters from spectra has been done in the literature using supervised learning methods, employing linear regression \citep{cannon}, neural networks \citep{starnet, astroNN}, or a combination of classic synthetic models of spectra and simple neural networks \citep{payne}.
However, typical supervised approach is inherently limited by the amount and quality of annotated data.
Data annotation is often tedious and time-consuming, and it is not always possible to obtain reliable labels \citep{stellar_find}.

``Unsupervised learning'' methods \citep[p.~14]{Murphy_22} are appealing due to their ability to discover inner structures or patterns in data without relying on labels.
These learned representations attempt to capture essential data features and can be beneficial in various tasks, such as clustering or reconstruction.
Although representations may capture the underlying patterns of the data, there is no guarantee that they align with our human understanding or are easily interpretable in specific contexts, such as spectral parameters \citep{stellar_find, Nima_2021}.

Autoencoders (AEs) are a key technique in unsupervised learning that focus on learning a low-dimensional representation of high-dimensional input data \citep[p.~675]{Murphy_22}.
AEs consist of two parts: an encoder and a decoder.
The encoding process involves passing the data through a ``bottleneck'' a middle part of the network where the data are transformed into a ``latent representation.''

To ensure that the latent representation is useful and the network does not learn merely an identity mapping, some form of regularization must be applied to the bottleneck.
In this paper, we achieve compression of the input by choosing a bottleneck that is much smaller than the input \citep[bottleneck autoencoder;][p.~667]{Murphy_22}.

The latent representation is crucial for data reconstruction and analysis, as it encodes essential underlying features for these tasks.
The space composed of all possible latent representations is called the ``latent space.'' 
The decoding process attempts to reconstruct the input from its latent representation, which is the learning objective of the AEs.
AEs determine the optimal latent space without any supervision, and the resulting latent space is optimal vis-à-vis the reconstruction of the input.
This process does not require any labeled data at any step.
AEs are used for denoising \citep[p.~677]{Murphy_22}, dimensionality reduction, and data compression \citep[p.~653]{Murphy_22}.

However, the latent space of AEs poses several challenges.
One key issue, as mentioned, is the lack of a guarantee to extract meaningful variations in the data \citep{AE_interpretation}.
This is informally known as interpretability since the meaningfulness of latent space is subjective with respect to the target application.
Additionally, the deterministic nature of AEs, where each input corresponds to a single point in the latent space, poses risks of overfitting and limits the model's generalization capabilities to new, unseen data \citep{VAE}.

The concept of ``disentangled representation'' is a possible formalization of interpretability.
Disentangled representation refers to the ability of a model to autonomously and distinctly represent the fundamental statistical factors of the data \citep{sober}.
In the context of this paper, statistical factors refer to various independent characteristics of celestial objects, such as their luminosity, temperature, chemical composition, or velocity.
A model that learns a disentangled representation can represent each of these astronomical factors independently.

For a model to perform well on new data (beyond interpolation), the decoder must correctly interpret the cause-and-effect relationships of individual factors on the resulting spectrum \citep{montero_2022}.
Determining causality from purely observational data is generally an unsolved problem that requires strong assumptions \citep[pp.~44-62]{petersElementsCausalInference2017}.
The process of learning causality can be simplified when we have access to physical simulations, which would allow us to model and test causal relationships more effectively \citep[pp.~118-120]{petersElementsCausalInference2017}.

Variational autoencoders \citep[VAEs;][]{VAE} represent a promising category of models capable of achieving a disentangled latent space.
This ability stems from the way VAEs conceptualize and manage the latent space. Unlike traditional AEs, which generate a single latent representation for each input (such as a spectrum), VAEs produce a distribution over the latent space for the same input.
We provide more details of the two basic VAEs variants in Sect.~\ref{sec:vaes}.

Variational autoencoders have some associated training challenges.
They require careful tuning of the hyperparameters to ensure the latent space is properly utilized and the model does not collapse to a trivial solution \citep[posterior collapse;][pp.~796--797]{Murphy_23}.
Other well-known issues are blurry reconstructions that are caused by over-regularized latent space \citep[pp.~787--788]{Murphy_23}. 

Independent studies have employed VAEs for HARPS \citep{HARPS} and SDSS \citep{SDSS_hack} spectra to self-learn an appropriate representation, thereby attempting to eliminate the need for annotation entirely \citep{portillo, Nima_2021}.
Although it was shown to be empirically possible to obtain some spectral parameters \citep{Nima_2021}, it is not clear how to obtain all of them or control which ones are obtained.

The large-scale study conducted by \cite{sober} proves that it is not feasible to achieve unsupervised disentanglement learning without making implicit assumptions that are influenced by ML models, data, and the training approach.
These assumptions, commonly referred to as ``inductive bias'' in the field of ML \citep{bias}, are challenging to manage due to their implicit nature.
Therefore, to successfully acquire disentangled representations in practical situations, it is crucial to obtain access to high quality data, establish suitable distributions that accurately model the underlying structure and complexity of the data, and use labels at least for the model validation \citep{disentangle_real}.

In light of these challenges with unsupervised learning, ``semi-supervised learning,'' which combines the advantages of both supervised and unsupervised learning, is a promising approach.
It is beneficial when the amount of labeled data are limited or labeling is costly, but the unlabeled data are abundant \citep[p.~634]{Murphy_22}.
Semi-supervision utilizes all data to overcome the lack of labels.
However, semi-supervised learning in astronomy can be challenging due to unbalanced datasets (where classes are unevenly represented, often with a significant imbalance in the number of instances per class), covariate shifts, and a lack of reliably labeled data \citep{RGZ_semi}.

Semi-supervised\footnote{The term ``supervised AEs'' may also be encountered, where the unsupervised nature is implicitly part of AEs.} AEs incorporate label prediction on top of reconstruction to improve their performance \citep{Le_2018}.
The same principle can be applied to VAEs \citep{semi_VAE} to obtain semi-supervised VAEs.
There are many variants; for example, \cite{Le_2018} uses a pair of decoders—one for input reconstruction and another for label prediction.
Alternatively, \cite{semi_VAE} (model M2) employs two encoders: one to provide an unsupervised bottleneck and the other to handle label prediction.
The semi-supervised approach allows us to jointly solve the label prediction and ML simulation tasks.

However, not all architectural choices lead to the same outcomes.
For instance, the approach of \cite{Le_2018} does not use predicted labels for reconstruction, meaning that the reconstruction error is not back-propagated to influence the predictions. In contrast, architectures that incorporate predicted labels during the reconstruction process allow the reconstruction error to inform and refine the label predictions through back-propagation, thus potentially improving prediction accuracy.

Concentrating only on label prediction, we considered simulation-based inference \citep[SBI;][]{SBI_summary}.
SBI utilizes existing simulations to infer parameters without relying on annotated data.
SBI methods either examine a grid of parameters \citep{SBI_summary} or iteratively improve initial estimates of parameters, such as in Markov chain Monte Carlo (MCMC) methods \citep{miller_2020}.
Modern approaches to SBI \citep{SBI_summary} employ machine learning techniques, like normalizing flows \citep[NFs;][]{NormalizingFlows2016}, for precise density estimation and to increase the speed.
However, training NFs or conditional NFs, is challenging due to the high dimensions of HARPS spectra.
Another issue is the significant differences between synthetic and observed data.


We prefer VAEs because they offer an efficient and flexible approach to modeling high-dimensional data, such as HARPS spectra, with a simpler and faster alternative to traditional SBI.
By leveraging semi-supervised VAEs, we can directly obtain posterior distributions when the variational distribution is flexible enough to match the target distribution, which is ideal for cases with high-dimensional observations, low-dimensional parameter spaces, and unimodal posteriors.
Additionally, VAEs have the capability to discover new parameters and integrate both known and unknown factors into simulations, enhancing their accuracy and adaptability.
This allows for quick sampling of candidate solutions and significantly reduces computational demands.
Moreover, VAEs can serve as a preprocessing step to reduce dimensionality, optimizing computational efficiency for downstream tasks, including those that use NFs for modeling more complex and multi-modal distributions.

We considered other generative models for ML simulation, such as diffusion models \citep{DiffusionModels}, NFs \citep{NormalizingFlows2016}, and generative adversarial networks \citep[GANs;][]{GAN}.
These powerful ML models project a vector of Gaussians into the simulations.
If desired, these models can condition the output on spectral parameters to provide greater control over the simulations.
Their flexibility allows them to adapt to different types of data and applications, providing a robust framework for generating realistic simulations across various domains.
However, these models have several limitations that make them less suitable for our specific ML simulation task.

These models intentionally function as black boxes, where the connection between the latent space and the generated simulation is hidden.
As a result, we obtain simulations with variations without understanding their origin.
This is acceptable in domains where explicit parametrization of the target domain is not possible or practical, such as in natural scenes for computer vision.
However, in spectroscopy applications, it is desirable to control ML simulations with known or discovered parameters.
Introducing variations into simulations merely to create an illusion of realism, without understanding the underlying processes, is meaningless in our context.
Additionally, training these models demands a large amount of computational resources to achieve good results.

A practical example of a GAN application is CYCLE-STARNET \citep{cycle_starnet}, which combines GANs and autoencoders to transform synthetic spectra into realistic ones and vice versa.
The method introduces two latent spaces: one shared between synthetic and real data and one specific to real observations.
By learning mappings between these spaces, CYCLE-STARNET can enhance the realism of synthetic spectra and facilitate the transformation of observed spectra back into the synthetic domain. The learning process is unsupervised, and the method does not provide label prediction.

In this study, we chose supervised VAEs for the ML simulation task because they facilitate explicit representation of stellar parameters.
VAEs can create a structured latent space that combines both known and unknown factors.
This allows us to obtain both the simulated spectrum and the parameters that were used.
Unlike other generative models, AEs and VAEs directly learn how to project the structured latent space to observations, making this approach deterministic and aligning with practical scientific requirements where we aim to minimize random variances in the output.
Furthermore, VAEs are significantly faster to train and can handle larger data dimensions than diffusion models or NFs.
Finally, as demonstrated in \cite{HighDimDiffusion2022}, compressing data with AEs is necessary to make diffusion models with high-dimensional data feasible.
This demonstrates that scalable compression techniques, such as AEs or VAEs, are still highly relevant for these models.

We enrich the training data with simulated data to create a balanced dataset with reliable labels.
We mix these synthetic data with real data to minimize the impact of covariate shifts.
We also explore splitting the latent space in a supervised (label-informed) and an unsupervised part, so as to ensure that the model learns the correct labels where possible, while utilizing disentangled methods for its unsupervised parts.

We identified a gap in the literature concerning the semantic correctness of ML simulations, specifically whether individual labels in these simulations behave according to theoretical expectations (causality).
In this study, we suggest modifications to existing methods and novel metrics to explore and solve this problem.

\subsection{Semi-supervised latent space}
Semi-supervised latent space $\mathcal{B}$ is a tool for representing input spectra that is influenced by both labels and the spectra themselves.
Within the latent representation $\mathbf{b} \in \mathcal{B}$, ``factors'' specifically denote the set of independent core characteristics or attributes that have been abstracted from higher-dimensional input spectra $\mathbf{s}$.
We further divide the latent representation into two components: ``label-informed factors'' $\hat{\mathbf{l}}^{\text{LIF}}$ and ``unknown factors'' $\mathbf{u}$, that is, $\mathbf{b} = (\hat{\mathbf{l}}^{\text{LIF}}, \mathbf{u})$.
The label-informed factors $\hat{\mathbf{l}}^{\text{LIF}}$ are supervised by known spectra parameters $\mathbf{l}$ from a catalog, where $\mathbf{l}$ undergoes a normalization process to ensure a consistent scale and distribution across the dataset.
This ensures that both the label-informed factors and the original labels are consistently scaled, enhancing the ability of our machine learning models to learn and generalize from the data effectively.
In this study, the symbol $\hat{\ }$ denotes prediction.

The normalization procedure adjusts label to follow a standard normal distribution. Specifically, the normalization and scaling of the $k$th label are defined as
\begin{align}
    \underline{l}^k = \frac{l^k - \mu^k}{\sigma^k} \quad \text{and} \quad l^k_{\text{scale}} = \frac{l^k}{\sigma^k}
    \label{eq:norm},
\end{align}
where $\mu^k$ and $\sigma^k$ are the mean and standard deviation of the $k$th label across the dataset, respectively.
The scaling operation enables the addition of unnormalized labels with normalized labels by first scaling the unnormalized labels.
In this study, $\underline{\mathbf{l}}$ denotes normalized labels and $\mathbf{l}_{\text{scale}}$ denotes scaled labels.

The unsupervised part, unknown factors $\mathbf{u}$, represents undetermined spectral parameters and other statistically relevant features.
Working in tandem with the supervised label-informed factors, these unknown factors help create a more comprehensive and informative representation of the latent space.
This holistic approach allows us to uncover hidden patterns and relationships within the spectral data, leading to more accurate ML simulation and label prediction as shown in Sect.~\ref{sec:results}.

We use the term ``label-aware'' when talking about models that utilize labels during the training process, regardless of whether the approach is fully supervised or semi-supervised.
Later, we assess the impact of the labels by comparing label-aware and unsupervised models.

\subsection{Inference tasks}
\label{sec:inference}
In ML, an inference task refers to a specific problem that we aim to solve.
This includes stating the assumed deployment condition, in other words, what data the model will receive and what output it should produce.
Specifying and formalizing the inference tasks is crucial for designing the ML models and selecting appropriate loss functions for learning the models.

We assume there is a dataset in the form $\{\mathbf{s}_i, \mathbf{l}_i, \mathbf{u}_i\}_{i=1}^D$.
Here, $D$ is the number of samples in the dataset, $\mathbf{s}_i \in \mathbb{R}^N$ represents an observed spectrum, $N$ is the number of pixels in a spectrum, $\mathbf{l}_i \in \mathbb{R}^K$ denotes the associated labels, where each label corresponds to a known spectral parameter such as temperature or airmass, $K$ denotes the number of labels, $\mathbf{u}_i \in \mathbb{R}^L$ denotes unknown factors, where each factor corresponds to an undetermined spectra parameter, and $L$ is the number of assumed unknown factors.
The wavelength vector is constant across all samples, and therefore, the spectrum $\mathbf{s}_i$ is represented just by the flux vector.

In this work, the spectra are standardized by dividing each spectrum by its median value, equivalent to the approach in \cite{Nima_2021}.
This process is formally defined as
\begin{equation}
\mathbf{s} \leftarrow \frac{\mathbf{s}}{\text{median}(\mathbf{s})},
\label{eq:spectra_norm}
\end{equation}
where $\text{median}(\mathbf{s})$ computes the median flux value across all pixels in the spectrum.
This step is fundamental for our ML models, ensuring that the spectral data are prepared consistently for all samples.
In this study, we always use the standardized spectra.

The true generative process $M$ is unknown and complex, involving both known and unknown factors.
We formalize this as a generative process $M: (\mathbf{l}, \mathbf{u}) \rightarrow \mathbf{s}$, which maps known and undetermined parameters to the observed spectrum.
Traditional simulations omit $\mathbf{u}$ and only map $\mathbf{l}$ to $\mathbf{s}$.

Our prime inference task aims to reverse the generative process $M$ and make label prediction $\hat{\mathbf{l}}$.
Our secondary inference task aims to model the generative process $M$, including the unknown factors.
We call our secondary task ML simulation, to better distinguish it from the traditional simulations approach using the applicable physical laws.
We seek $M$ such that if they obtain latent representation $\mathbf{b} = (\underline{\mathbf{l}}, \mathbf{u})$ for a particular spectrum $\mathbf{s}$, intervening on a factor in $\mathbf{b}$ should result in changes to $\mathbf{\hat{s}}$ according to the semantics of that factor.
For example, modifying a factor associated with radial velocity should produce a Doppler shift in the stellar lines and no other effect.

\subsection{Learning problem statement}
Once the inference tasks have been defined, we can formalize the learning problem.
Learning is the process of obtaining a model that solves the defined inference tasks.
This involves selecting appropriate ML models, defining a loss function that penalizes discrepancies between the model outputs and the actual observations, and partitioning the data to facilitate training, validation, and testing of the model.
An ML model is completely described by its parameters and hyperparameters.
For neural networks, ML hyperparameters--such as architecture, learning rate, batch size, and optimization methods--externally characterize the model and are either not directly related to the loss function or we want to keep them constant during training.
ML parameters, which include the weights and biases within the model, determine the neural network's output and directly influence the loss function.

During training, the ML parameters are fitted to the training data using backpropagation \citep[p. 434]{Murphy_22}, which computes the gradient of the loss function with respect to the ML parameters.
This gradient is then used for non-linear optimization.
Since ML hyperparameters do not contribute to the loss function gradient, we have to experiment with different combinations of ML hyperparameters to optimize the model's performance.
During ``model selection,'' we train a model for each combination of ML hyperparameters and validate each model by evaluating the loss function (or some other metric) on the validation data.
We select the model with the best performance among all other models. 
Optionally, during testing, we evaluate the selected model on the testing data to obtain an unbiased estimate of the model's expected performance.

We aim to train ML models that approximate $M$ and $M^{-1}$ using the dataset $\{\mathbf{s}_i, \mathbf{l}_i, \mathbf{u}_i\}_{i=1}^D$.
We have chosen the encoder-decoder architecture \cite{ae_rev} as the suitable ML model.
Encoders map high-dimensional input (spectra) to low-dimensional output (labels), effectively approximating $M^{-1}$.
Decoders, conversely, map low-dimensional input (labels) back to high-dimensional output (spectra), serving as a suitable approximation of $M$.

We train the encoder model either in isolation using a supervised approach, or in conjunction with the decoder using a semi-supervised or unsupervised approach.
Semi-supervised learning enables us to leverage labeled and unlabeled data, potentially enhancing performance.
Moreover, the secondary inference objective of the ML simulation cannot be optimally achieved through supervised learning alone, since, by definition, we cannot supervise unknown factors $\mathbf{u}$.

All of our models minimize the loss functions that represent the disparity between the model output and the actual observations.
Next, we briefly discuss progressively more complex models for the
inference tasks and the corresponding learning loss functions.

\subsection{Machine learning models}
\label{sec:ml_models}
Here, we present the ML models that target our inference tasks as described in Sect.~\ref{sec:inference}.
We begin with encoders and decoders as they are the foundational elements of the subsequent models.
Beyond serving as foundational elements, an individual decoder can be used for ML simulation tasks, whereas an individual encoder is useful for label prediction.
Building upon these, we develop both semi-supervised and unsupervised AEs.
This section concludes with a description of semi-supervised VAEs and their derived methods.
All AE-based methods are capable of jointly addressing label prediction and ML simulation tasks.
The source code, including model implementations, data preprocessing, and training scripts, can be accessed at\footnote{\url{https://github.com/cvrcek/HARPS-ML-Spectra}}.

\subsubsection{Encoders}
\label{sec:supervised}
Encoders are part of AEs and the basic model that can provide label prediction.
We employ convolutional neural networks (CNNs) that stack layers of convolutions along with differentiable non-linear activation functions to process an input spectrum. The CNN output is processed by a single fully connected layer to learn the encoding $q^c_\phi$ from spectrum $\mathbf{s}$ to latent representation $\mathbf{b}$.
The CNN architecture is based on the encoder proposed in \cite{Nima_2021}, and a detailed description is given in Table~\ref{tab:encoder_CNN}.
We can achieve label prediction by ignoring $\mathbf{u}$ ($\mathbf{u} = \emptyset$).

We defined the loss function for label prediction using encoders as the mean absolute label difference:
\begin{equation}
L_{\text{lab}}(\phi) = \mathbb{E}_{(\mathbf{s}, \mathbf{l}) \sim p_\mathcal{D}}
\left[
\frac{1}{K} \sum^{K}_{k=1} \left| \underline{l}^k - q^{c,k}_{\phi}(\mathbf{s}) \right|
\right],
\label{eq:lab}
\end{equation}
where $\mathbb{E}$ is expected value, $l^k$ is the $k$th label (ground truth), $q^{c, k}_{\phi}(\mathbf{s})$ is the predicted value for $k$th label, $p_\mathcal{D}$ is the distribution of the spectra $\mathbf{s}$ and associated catalog values $\mathbf{l}$, and $K$ is the number of labels for each spectrum.
For instance, in the case of HARPS observational spectra, the distribution is empirical, based on the actual data samples we have. We sample from this distribution by randomly selecting a spectrum.

The effectiveness of training encoder in isolation depends on the availability of learning data with sufficient variability in all factors.
However, in our data set, only 1,498 unique spectra are fully annotated (as described in Sect.~\ref{sec:experiments}).
Most samples have incomplete annotations, and many spectra lack annotations altogether.
Furthermore, corrupted, incorrect, or mislabeled data can negatively impact the supervised learning process.

Therefore, we investigate the incorporation of unsupervised methods to improve the accuracy of label prediction. A key ingredient in learning directly from spectra is the ability to reverse the encoding process through the use of decoders.

\subsubsection{Decoders}
\label{sec:simulation}
Decoders are part of AEs and the basic model that provides ML simulation.
In our decoder architecture $p_\theta$, we have used residual network \citep[ResNet;][]{resnet}, as shown in Table~\ref{tab:decoder_resnet}, or CNN, as illustrated in Table~\ref{tab:decoder_CNN}
In either case, our objective is to learn the ML parameters $\theta$.
The loss function for the ML simulation is the mean reconstruction error:
\begin{equation}
L_{\text{sim}}(\theta) =
 \mathbb{E}_{(\mathbf{s}, \mathbf{l}) \sim p_\mathcal{D}}
\left[
    \frac{1}{N} \sum^N_j \left| s^{j} - p^j_{\theta}(\underline{\mathbf{l}})\right| 
\right],
\label{eq:sim}
\end{equation}
where $s^{j}$ represents the  flux at pixel $j$ of spectrum $\mathbf{s}$, $p^j_{\theta}(\underline{\mathbf{l}})$ is the predicted flux at pixel $j$ for labels $\mathbf{l}$ and $N$ is the total number of pixels in the spectrum.

A potential downside of CNN architectures is that each layer processes only the output of its preceding layer.
While ResNet mitigates this by incorporating short-range skip connections across two layers, this strategy may still be suboptimal.
As data pass through deeper layers, the input labels or bottleneck representations may become diluted, leading to a loss of important information.
To address this, we considered alternative architectures, including skip-VAEs \citep{skipVAE}, DenseNet \citep{DenseNet}, and FiLM layers \citep{FiLM_layer}, which can allow labels to influence any layer.
Nevertheless, in this work, we have focused on CNN and ResNet architectures, deferring exploration of these alternatives to future research.

\subsubsection{Autoencoders and downstream learning}
\label{sec:AE}
Autoencoders are a type of machine learning architecture that enables unsupervised learning directly from data.
An AE is constructed by connecting an encoder to a decoder.
It does not require labels because it feeds the encoder's output directly into the decoder to achieve reconstruction.
AEs are trained to minimize the reconstruction loss between the input and the reconstructed output.

Reconstruction loss is a measure of fidelity, quantifying how closely the reconstructed output matches the input.
It also reflects the efficiency of the bottleneck, ensuring that it retains essential features for reconstruction.
The reconstruction loss is again defined as the mean absolute difference across all pixels:
\begin{equation}
L_{\text{rec}}(\theta, \phi) =
 \mathbb{E}_{\mathbf{s} \sim p_\mathcal{D}}
\left[
    \frac{1}{N} \sum^N_j \left| s^{j} - p^j_{\theta}(q^c_\phi(\mathbf{s}))\right| 
\right].
\label{eq:rec}
\end{equation}
Here, $p^j_{\theta}(q^c_\phi(\mathbf{s}))$ is the AE's prediction for the same pixel.
Optimizing this loss over $(\theta, \phi)$, the AE is trained to effectively capture the salient features of the spectral data necessary for reconstruction.
The resulting latent representation might be practical for new tasks, such as label prediction, since it is much easier to process  small latent representation instead of the full spectrum.

Secondary tasks that utilize compressed representations from an AE are called downstream tasks.
The typical deployment scenario occurs when we have abundant high-dimensional unlabeled data, of which only a small subset is labeled, and we aim to predict the labels. The workflow consists of two steps: first, learning an AE using the unlabeled data. This allows us to map high-dimensional data to a lower-dimensional space. Second, we use the low-dimensional representations (such as spectra) from the labeled subset to learn label prediction. There is no guarantee that an AE will learn a latent space useful for the target downstream task. This is influenced by the choice of the AE's architecture, optimization techniques, and the properties of the data and labels. Therefore, the downstream task, which is usually the main objective, can serve as a criterion for model selection.

Our downstream tasks are label prediction and ML simulation.
We chose linear regression for label prediction for two reasons.
First, we are interested in AEs that provide a meaningful, disentangled latent space that can be straightforwardly translated into labels.
A more complex model might recover labels from an entangled latent space, which has no clear connection to the labels.
Second, since our second inference task is ML simulation, we need to map the labels back to the latent representation, which is then mapped to a spectrum.
Complex models with multiple layers and non-linearities would be challenging to invert.

We used the following linear regression to map latent representation $\mathbf{b}$ to labels $l$:
\begin{equation}
    \mathbf{\hat{l}} = \sum_{k=1}^B {w}^k \mathbf{b}^k + w^k,
\end{equation}
where $\mathbf{w}$ are weights defining the linear regression, $\mathbf{\hat{l}}$ are predicted labels, and $B$ is the size of the bottleneck.
The model was trained using ordinary least squares linear regression.

The downstream learning represents a two-step approach that links unsupervised preprocessing and supervised learning for label prediction or ML simulation.
Next, we investigate the semi-supervised methodology where we use a single training phase that combines spectra and labels.

\subsubsection{Semi-supervised autoencoder}
\label{sec:semi_supervised}
We can trivially add supervision to the unsupervised AE, thus obtaining semi-supervised AE.
Similarly to the supervision in Sect.~\ref{sec:supervised}, we use the known labels $\mathbf{l}$ to supervise the label-informed factors $\hat{\mathbf{l}}^{\text{LIF}}$.
Hence, we have label-informed factors that are influenced by both unsupervised and supervised objectives.
In addition, we allow the unknown factors $\mathbf{u}$ to learn statistically meaningful information not provided by the known labels $\mathbf{l}$. 
We achieve this by expanding the loss function in Eq.~\eqref{eq:rec} while preserving the architecture of unsupervised AEs.

The integration of unsupervised and supervised learning objectives is captured by the following loss function, which combines the reconstruction loss $L_{\text{rec}}$ from Eq.~\eqref{eq:rec} for unsupervised learning with the label loss $L_{\text{lab}}$ from Eq.~\eqref{eq:lab} from supervised learning:
\begin{equation}
L_{\text{AE}}(\theta, \phi) = 
L_{\text{rec}}(\theta, \phi) + \lambda_{\text{lab}} L_{\text{lab}}(\phi),
\label{eq:AE}
\end{equation}
where $\lambda_{\text{lab}}$ is a hyperparameter that allows balancing between reconstruction and label loss.

By design, the supervised portion $\hat{\mathbf{l}}^{\text{LIF}}$ of the latent representation $\mathbf{b}$ is meaningful and disentangled.
However, the unsupervised portion $\mathbf{u}$ faces the same problems as latent representation in AEs: a lack of interpretability and entanglement.
As a result, unsupervised nodes can become entangled with supervised nodes in the bottleneck. This is especially problematic for ML simulations, as the unsupervised nodes can interfere with the role of the supervised nodes during the simulation.

As a solution, we investigated methods to regularize the bottleneck to achieve disentanglement of the unsupervised nodes.
By imposing suitable constraints on the bottleneck, the model can maintain the integrity of the supervised nodes while ensuring that the unsupervised nodes capture information not contained in the known labels.
Additionally, disentangling the unsupervised nodes allows them to be integrated into the ML simulation because we can easily sample from independent unsupervised nodes.

\subsubsection{Semi-supervised variational autoencoder}
\label{sec:infoVAE}
Variational autoencoders are suitable for learning disentangled latent representations.
The VAE is a probabilistic variant of AEs adept at learning disentangled latent representations.
The key difference between VAEs and AEs lies in treating the latent representation as a distribution over the latent space, versus a single latent representation.
We employ the $\beta$-VAE, a modification of the classic VAE.
This adaptation introduces a ML hyperparameter $\lambda_{\mathbb{KL}}$ that enables a flexible balance between reconstruction quality and the penalization of the latent representation distribution, thus facilitating a more controlled disentanglement of features.

The loss function of the $\beta$-VAE is
\begin{equation}
L_{\beta\text{-VAE (U)}}(\theta, \phi, \lambda_{\mathbb{KL}}) = 
L_{\text{rec}}(\theta, \phi)  - 
\lambda_{\mathbb{KL}}
\mathbb{E}_{\mathbf{s} \sim p_\mathcal{D}}
\Bigl[
D_{\mathbb{KL}}(q^p_\phi(\mathbf{b} \mid \mathbf{s}) || p_b(\mathbf{b}))
\Bigr],
\label{eq:BVAE_original},
\end{equation}
where $p_\mathcal{D}$ is a data distribution,
$q^p_\phi$ is a probabilistic encoder that maps spectra $\mathbf{s}$ to a distribution over latent space $\mathbf{b} \sim q^{p}_\phi(\cdot \mid \mathbf{s})$,
$p_b$ is the spherical normal distribution that represents the implicitly disentangled prior over the latent space,
$\lambda_{\mathbb{KL}}$ is the weight placed on the Kullback-Leibler (KL) term (called $\beta$ in \cite{Beta_VAE}), and $D_{\mathbb{KL}}$ is the KL divergence described in Eq.~\eqref{eq:KL}.
The equation reflects how the $\beta$-VAE's objective function, derived from the original VAE, balances reconstruction accuracy and the latent space's regularization.
Further insights into the original $\beta$-VAE objective and its relation to our implementation can be found in Appendix~\ref{sec:VAE_basics}.

To achieve supervision, we added the label loss to the objective in Eq.~\eqref{eq:BVAE_original}:
\begin{align}
L_{\beta\text{-VAE}}(\theta, \phi, \lambda_{\mathbb{KL}}) &= 
L_{\beta\text{-VAE (U)}}(\theta, \phi, \lambda_{\mathbb{KL}}) + \nonumber \\ 
& \lambda_\text{lab} \mathbb{E}_{(\mathbf{s}, \mathbf{l}) \sim p_\mathcal{D}}
\left[
\frac{1}{K} \sum^{K}_{k=1} \left| \underline{l}_{k} - b_{k}(\mathbf{s}) \right|
\right],
\label{eq:BVAE}
\end{align}
where  $\mathbf{b}(\mathbf{s}) \sim q^p_\phi(\cdot \mid \mathbf{s})$ denotes a vector sampled from $ q^p_\phi$ given the input $\mathbf{s}$, and $b_k(\mathbf{s})$ represents the $k$-th element of that vector.
This sampling strategy incorporates the stochastic nature of $q^p_\phi$ into the supervised learning framework.

An inherent challenge associated with the $\beta$-VAE framework is posterior collapse, where the latent representation fails to capture meaningful information from the input \citep[pp.~796--797]{Murphy_23}.
This results in underutilization of nodes, effectively reducing the size of the bottleneck.

To address the problem of posterior collapse in $\beta$-VAEs, we adopt a mutual information-based method known as the Informational Maximizing Variational Autoencoder \citep[InfoVAE;][]{infoVAE}.
This approach balances the disentanglement and informativeness of the latent representation.
The corresponding loss function for unsupervised InfoVAE (U) is defined as
\begin{align}
    L_{\text {InfoVAE (U)}}(\theta, \phi, \lambda_{\text{MI}}, &\lambda_\text{MMD}) = 
    L_{\text{rec}}(\theta, \phi)
    - \nonumber \\
    & (1-\lambda_{\text{MI}}) \mathbb{E}_{\mathbf{s} \sim p_{\mathcal{D}}} D_{\mathbb{KL}}\left(q^p_\phi(\mathbf{b} \mid \mathbf{s}) \| p_b(\mathbf{b})\right)- \nonumber \\
    & (\lambda_{\text{MI}}+\lambda_\text{MMD}-1) D_{\mathbb{KL}}\left(q^p_\phi(\mathbf{b}) \| p_b(\mathbf{b})\right),
    \label{eq:infoVAE_equivalent},
\end{align}
where a higher $\lambda_{\text{MI}}$ means $\mathbf{s}$ and $\mathbf{b}$ have higher mutual information, and higher $\lambda_\text{MMD}$ brings $q^p_\phi(\mathbf{b})$ closer to the priors $p_b(\mathbf{b})$.

The term $1 - \lambda_{\text{MI}}$ serves a purpose similar to $\lambda_{\mathbb{KL}}$ in the original $\beta$-VAE model.
The primary distinction between $\beta$-VAE (U) and InfoVAE (U) lies in the term $D_{\mathbb{KL}}\left(q^p_\phi(\mathbf{b}) | p_b(\mathbf{b})\right)$ and the addition of extra ML hyperparameter.
InfoVAE (U) allows us to balance disentanglement, reconstruction, and informativeness.
For a more in-depth understanding and computational specifics of the InfoVAE loss function, please refer to Appendix~\ref{sec:infoVAE_basics}.

By adding the label loss, we obtain the loss function for supervised InfoVAE:
\begin{align}
    L_{\text {InfoVAE}}(\theta, \phi, \lambda_{\text{MI}}, \lambda_\text{MMD}) =
    &L_{\text {InfoVAE (U)}}(\theta, \phi, \lambda_{\text{MI}}, \lambda_\text{MMD}) + \nonumber \\
    & \lambda_\text{lab} \mathbb{E}_{(\mathbf{s}, \mathbf{l}) \sim p_\mathcal{D}}
    \left[
    \frac{1}{K} \sum^{K}_{k=1} | \underline{l}_{k} - b_{k}(\mathbf{s}) |
    \right].
    \label{eq:infoVAE}.
\end{align}
This final formulation, InfoVAE, integrates the strengths of InfoVAE with supervised learning elements.
The complete visualization is shown in Fig.~\ref{fig:illustration}.

\section{Application of machine learning to spectra}
\label{sec:experiments}
In this section we describe the results of applying the methods introduced in Sect.~\ref{sec:methods} to the HARPS spectra described below in Sect.~\ref{sec:data}.
We used a simulated dataset to cover various combinations of spectral parameters.
These simulations are obtained from synthetic spectral energy distributions (SEDs) computed from the Kurucz~\citep{kurucz} stellar atmosphere models, processed through the instrument's Exposure Time Calculator \citep[ETC;][]{ETC2}.
The ETC simulates atmospheric effects and those of the measurement apparatus, including the telescope and the HARPS instrument itself.
To further increase the variability of the data, we propose transformation strategies to generate new data on the fly from the simulated data. This approach, known as data augmentation in ML terminology, is typically used to improve training by effectively increasing the quantity of data.
These datasets aim to encompass a comprehensive range of astrophysical scenarios.

Next, we describe our approach to train and optimize the models.
Then, we present the metrics used to evaluate the quality of label prediction.
We evaluate the quality of ML simulation using three metrics, each targeting different properties.
We use the standard reconstruction error to investigate the faithfulness of the reconstruction.
Our two generative metrics are designed to measure how well an ML model grasps the cause-and-effect relationship between the spectral parameters and the output spectrum.
Finally, we discuss our approach to model selection.

\subsection{Data}
\label{sec:data}
In this section we describe both the use of real spectra from the HARPS instrument and our methodology for generating simulated spectra to enrich our dataset.
This dual approach broadens our capabilities for both training and evaluating our models.
The summary of the datasets we are using is in Table~\ref{tab:datasets}.

The selected dataset is comprised of HARPS observations ranging from October 24, 2003, to March 12, 2020 (the instrument is still in active operation, so more observations are being added regularly).
Through the ESO Science Archive, we have access to 267,361 fully reduced 1-dimensional (1D) HARPS spectra, consisting of flux as a function of wavelength.
The processing that went from raw data to products is described in the online documentation.\footnote{ \url{https://www.eso.org/rm/api/v1/public/releaseDescriptions/72}} 
About 2\% of the raw science files failed processing into 1D spectra and are therefore not included in our analysis.
Each spectrum has a spectral resolving power of $R = \lambda/\Delta \lambda\simeq115,000$, and covers a spectral range from 380 to 690~nm, with a 3~nm gap in the middle.

Our machine learning experiments necessitate verified labels, which we consider the ``ground truth.''
We have two sources for the physical parameters used as labels: (1) the TESS Input Catalog (TIC, \cite{TESS}), as already used by \cite{Nima_2021}, serves as the source of our labels for the real data; and (2) the set of spectral parameters used to generate the ETC dataset. ETC labels are the most reliable with absolute control and consistency.

\begin{figure}[htbp]
    \centering
    \includegraphics[width=0.5\textwidth]{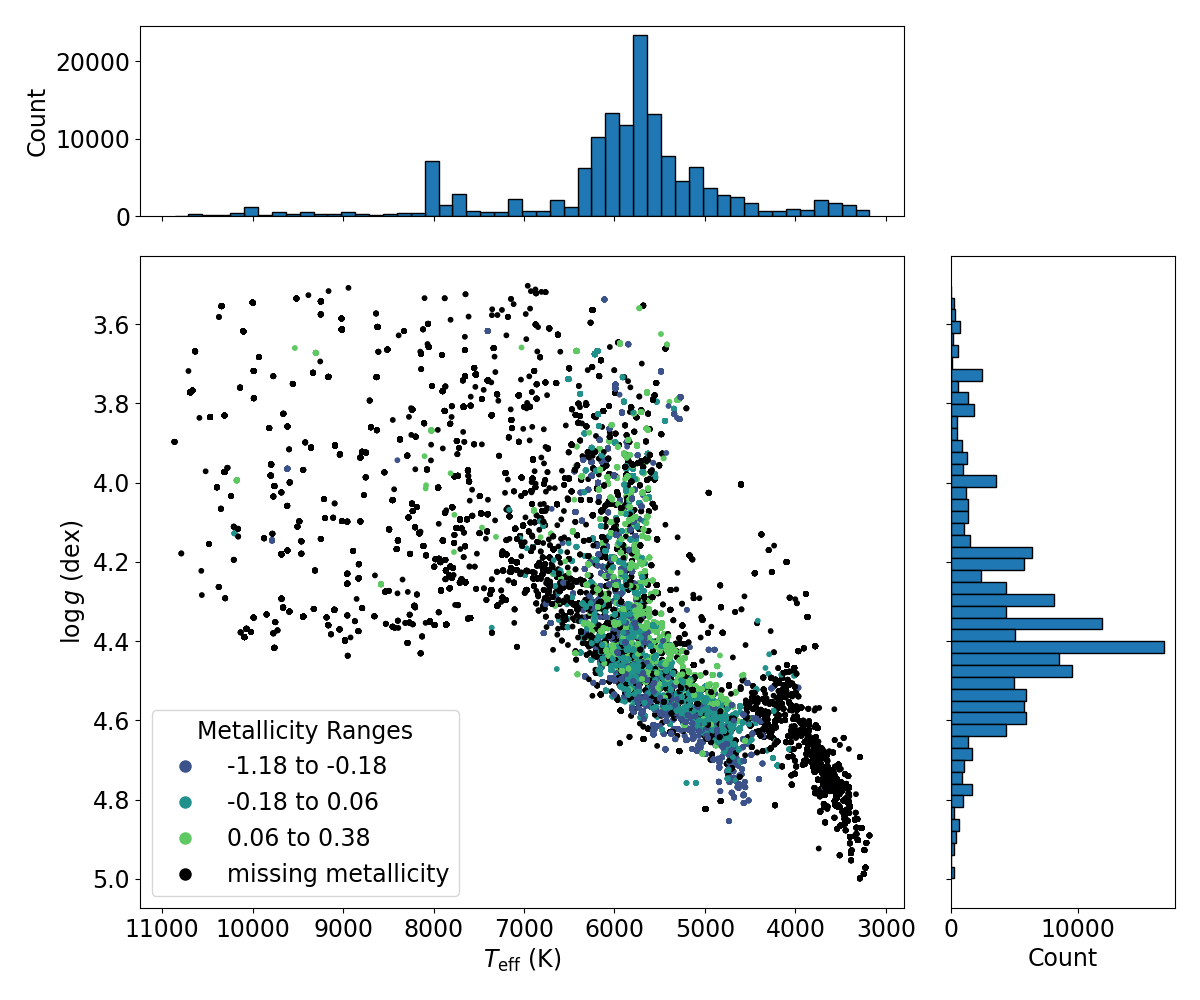}
    \caption{
        Distribution of effective temperature, surface gravity, and metallicity in the HARPS dataset.
    }
    \label{fig:teff_logg_plot}
\end{figure}

The TESS Input Catalog (and HARPS dataset) exhibits bias that is shown in Fig.~\ref{fig:teff_logg_plot}, where each point represents a single observation.
The data in this figure has already been filtered to remove underrepresented labels, as described below.
The top histogram shows the distribution of the effective temperature, while the right histogram shows the distribution of the surface gravity.
The color of the points represents metallicity, with black points indicating missing metallicity data.
The figure shows that the main sequence is well represented, while other regions are underrepresented.

Therefore, we have filtered out spectra with labels that are significantly underrepresented in the catalog.
Specifically, we have removed spectra with temperatures below 3000~K and above 11000~K, metallicities below -1.2~dex and above 0.4~dex, and surface gravities below 3.5~dex and above 5~dex.
This filtering was a practical necessity for model training since these ranges contain less than 2\% of the samples, making it difficult to achieve reliable training, validation, and testing. 
This limits the model's ability to generalize to these edge cases. We intend to address these limitation  in future work through proper uncertainty quantification methods that would allow the model to express reduced confidence when making predictions near or beyond these boundaries.

We prepared the simulated data in two stages.
First, we generated the intrinsic Spectral Energy Distribution emerging from the stars ("SED dataset") by employing the ATLAS9 software \citep{kurucz}, facilitated by the use of Autokur \citep{Autokur} to generate a dense grid that samples the stellar parameter space covered by the real data.
This is driven by the following stellar parameters: effective temperature, surface gravity, and the chemical composition of the star.
We fixed the microturbulent velocity at $2 \si{\kilo\meter\per\second}$, because this number is rarely available in catalogs for an individual spectrum.

To generate stellar parameters for our SED dataset, we sampled from uniform distributions with the following ranges: effective temperature from 3000~K to 11000~K, surface gravity from 3.5~dex to 5.0~dex, and metallicity from -1.2~dex to 0.4. This approach ensures that our simulated data cover the same parameter space as the filtered HARPS dataset, allowing for direct comparisons while also providing more uniform coverage of underrepresented regions.

In the final step, we used the ETC \citep{ETC2} to create a simulated dataset that more closely resembles the real HARPS dataset by including in the SEDs generated above the effects of the observing process, such as the imprints of the telescope, the instrument, and the Earth's atmosphere. The spectral parameters are the magnitude, atmospheric water vapor, airmass, fractional lunar illumination, seeing, and exposure time.
When referring to the ETC data in the text, we implicitly mean the combination of SED and ETC data. Full technical details are provided in Sect.~\ref{sec:ETC}.
In total, we generated 44,000 simulated HARPS spectra.

We utilize ETC data for pre-training, regularization, and transfer learning.
We can generate an almost arbitrary combination of spectral parameters in the desired quantity.
Although having more data is generally beneficial, combining data from multiple sources can be challenging and does not guarantee positive outcomes.
Therefore, we investigate the impact of mixing ETC and HARPS data on model performance through several experiments, specifically examining the effect of adding simulated data to the real one.

The labels are further categorized as ``intrinsic'' (temperature, metallicity, and surface gravity) and ``extrinsic'' (radial velocity, airmass, and barycentric Earth radial velocity-BERV).
The overview of labels availability for HARPS is presented in Table~\ref{tab:labels_counts}.

Each model can use either the observational HARPS dataset, the ETC dataset, or a mixture of both.
We split the HARPS dataset into $90\%$ data for training, $5\%$ data for validation, and $5\%$ for testing.
Furthermore, some of our experiments utilized only 1\% of the available labels for the real HARPS dataset.
This is done to test and quantify how the accuracy and fidelity of the label prediction depend on the availability of input labels, which may be scarce in real-life scenarios.

For the ETC dataset, we generated separate datasets for training (42,000 samples), validation (1,000 samples), and testing (1,000 samples).
The labels were sampled independently, each chosen randomly from a uniform distribution.

To assess the generative properties of models, we construct a generative ETC dataset.
In this context, a generative dataset is one specifically designed to understand how isolated changes in certain labels affect the behavior and output of ML models.
The dataset is organized into subsets, with each subset dedicated to exploring the variations of a single specific label.

The assembly process includes several steps.
First, we generate a set of $c$ core samples.
Each core sample acts as a baseline configuration where all labels are set to uniformly sampled random values.
Next, for each core sample, we systematically alter one of $f$ labels and create $v$ variations.
In these variations, only the target label is altered from its baseline value in the core sample, while the other labels remain unchanged.

As a result of this process, the final dataset contains $c\cdot f\cdot v$ samples, representing an exploration of how changes in each label affect the generative properties of the models.
This dataset provides the basis for a detailed evaluation of the individual impact of each label.

\begin{table}
\caption{HARPS label availability.} 
\small
\begin{tabular}{lrrr}
    \toprule
    Label & Available & \makecell{Available for\\ unique target} & \makecell{Labels in catalog \\(percentage)}  \\
    \midrule
    $T_\text{eff}$ & 168728 & 4588 & 63.1\% \\
    $\log g$ & 145372 & 3582 & 54.4\% \\
    \text{[M/H]} & 120440 & 2108 & 45.0\% \\
    radvel & 142970 & 4469 & 53.5\% \\
    airmass & 267361 & 7274 & 100\% \\
    BERV & 267361 & 7274 & 100\% \\
    \hline
    All Labels & 76948 & 1498 & 28.8\% \\
    \bottomrule
\end{tabular}
\tablefoot{Available labels for 267,361 HARPS spectra, comprising 7,274 unique targets.
``Available'' is the total number of labels available.
``Available for unique target'' is the number of unique targets with labels available.
``Labels in catalog (percentage)'' refers to the portion of labels that are in the catalog, expressed as a percentage of the total number of spectra.
``All Labels'' is the number of targets with all labels available.}
\label{tab:labels_counts}
\end{table}
\normalsize

\begin{table*}[ht]
    \centering
    \caption{Datasets used.}
    \label{tab:datasets}
    \small
    \begin{tabular}{|p{4cm}|p{4.0cm}|p{4.0cm}|p{4.0cm}|}
    \hline
    & HARPS Dataset & ETC Dataset & Generative ETC Dataset \\
    \hline
    data acquisition & Oct 24, 2003 - Mar 12, 2020 & simulated using ATLAS9, Autokur, ETC & same as ETC \\
    \hline
    total spectra & 267,361 & 44,000 & 1960\\
    \hline
    spectral resolving power & $R \approx 115,000$ & matched to HARPS &  same as ETC \\
    \hline
    wavelength range & 380 - 690 nm (3 nm gap) & matched to HARPS & same as ETC \\
    \hline
    training - validation - testing split & 90\%-5\%-5\% & 42,000-1,000-1,000 & 0-980-980 \\
    \hline
    data augmentation & none & radial velocity \& $\mathbf{u}$ penalization & radial velocity \\
    \hline    
    utilization & real and mixed dataset; training, validation and testing & ETC and mixed dataset; training and testing & computation of GIS from Eq.~\ref{eq:gis} \\
    \hline
    \end{tabular}
\end{table*}

\subsection{Data augmentation}
\label{sec:augmentation}
As detailed in Sect.~\ref{sec:ETC}, our data collection method allows for the generation of synthetic spectra with arbitrary radial velocities without further reliance on the ETC tool.
This allows us to randomly alter radial velocities during training with minimal impact on performance, a technique known as ``data augmentation'' in ML.
Our augmentation reduces the likelihood of memorization, since no single spectrum is repeated with the same radial velocity.
Therefore, the augmentation encourages the model to disentangle the radial velocity more effectively.
For this study, we uniformly sample radial velocities between $-100$ and $100$ km/s.
However, this method does have drawbacks, including slower I/O and increased memory demands per individual spectrum.

Our second strategy for augmentation focuses on annotation.
We know that ETC spectra can be fully described by $\mathbf{l}$, and $\mathbf{u}$ is relevant only for real spectra.
We complied with this by allowing ETC spectra to utilize $\mathbf{u}$, but penalize its usage to encourage the model to prefer representations that do not rely on $\mathbf{u}$ for ETC data.
This penalization is implemented through a supervised loss, where we supervise $\mathbf{u}$ to be equal to the zero vector.

As a consequence, the ML model uses $\mathbf{u}$ only for real data.
This might help with disentanglement as any attempt by the ML model to entangle unsupervised nodes with supervised will be penalized for ETC data.
Furthermore, this strategy calibrates $\mathbf{u}$ as zero is connected to simulation data, while values different from zero inform about deviations from the simulated data.

Both strategies target the simulated data.
The first strategy is specific to radial velocity and increases the variability of the simulated data without impacting the real data, that is, we could achieve the same effect by simply sampling more simulated data.
The second strategy involves weak supervision of the unsupervised nodes $\mathbf{u}$ for simulated data during training. Consequently, the treatment of the real data is affected as well.
Therefore, the second strategy is more general.

\subsection{Model training and optimization}
In this section we describe the process of implementing and training the various models.
The models we focus on include stand-alone encoders, stand-alone decoders, AEs, $\beta$-VAEs, and infoVAEs, as detailed in Sect.~\ref{sec:methods}.
All models that include an encoder use the CNN encoder as specified in Table~\ref{tab:encoder_CNN}.
The decoding process is implemented by a CNN, detailed in Table~\ref{tab:decoder_CNN}, or a ResNet decoder as outlined in Table~\ref{tab:decoder_resnet}.
All models were implemented and trained using the PyTorch Lightning framework~\citep{Falcon_PyTorch_Lightning_2019}.

We aim to select the highest possible value for the learning rate, as it speeds up the training.
However, a learning rate that is too high results in spectral parameter divergence during training due to the vanishing or exploding gradient \citep[p.~443]{Murphy_22}.
We set the learning rate to $10^{-4}$ for models that use CNN in encoder or decoder configurations. 
A higher learning rate of $10^{-3}$ is acceptable for models exclusively based on ResNet due to their enhanced stability.
Thus, we can learn ResNet models significantly faster.

The training phase utilizes an Adam optimizer \citep{adam}, a stochastic gradient descent variant, to efficiently manage backpropagation \citep[p.~434]{Murphy_22} and update of ML parameters.
Informally, the Adam optimizer can be seen to be adaptively modifying the learning rate based on the optimization process of the ML parameters.
This optimizer includes a pair of ML hyperparameters, set to $(\beta_1, \beta_2) = (0.9, 0.999)$.
These ML hyperparameters help balance the influence of recent and past gradients. For more details on the Adam optimizer, see \cite{adam}.
The Adam optimizer is a common and often default choice. Since we did not encounter any problems, exploring further alternatives was not worthwhile.

The standard practice in ML training is to monitor a metric that is evaluated on the validation dataset and stop training once the metric starts increasing; this practice is called  ``early stopping.''
We initially chose the label prediction error as our stopping metric due to its relevance and quick evaluation. However, despite many days of training, we never observed the label prediction error increase; instead, it oscillated randomly while continuing to converge, resembling the double descent phenomenon for training epochs \citep{nakkiranDeepDoubleDescent2019}.
This unpredictable oscillation made setting an early stopping rule difficult.

As a result, we decided to train each model for a fixed 1,000 epochs, where a single epoch is a complete pass through the dataset. 
An epoch is composed of numerous mini-batches, with the batch size indicating the number of spectra processed in each mini-batch.
Given the extensive scale of our models and data, we opted for relatively small batch size, setting it to 32 spectra per batch.
For additional details on optimizers and general machine learning concepts, readers may refer to \cite{goodfellow2016deep,Murphy_22}.

\subsection{Label prediction error}
Our primary objective is focused on label prediction.
We aim to produce results that closely match the labels provided by the catalog, for which we need to assess how successful our ML models are.
The most straightforward approach is plotting the distribution of the errors.
For visualization of error distributions, we use kernel density estimation (KDE) with bandwidth set according to Scott's rule \citep{Scott_rule}, as shown in Fig.~\ref{fig:labels_detail}.
This visualization is useful for qualitative analysis but does not provide a single numerical value that summarizes the model's performance for the label prediction.

We achieved this by using the mean absolute error (MAE) between the predicted and the unnormalized ground truth labels:
\begin{equation}
\text{MAE}(\{\mathbf{s}_i, \mathbf{l}_i, \hat{\mathbf{l}}_i\}_{i=1}^{D}, k) = \frac{1}{D} \sum_{i=1}^{D} \left| \mathbf{l}_i^k - \hat{\mathbf{l}}_i^k \right|.
\label{eq:MAE}
\end{equation}
The set $\{\mathbf{s}_i, \mathbf{l}_i \}_{i=1}^{D}$ denotes the testing dataset, where $D$ is its size.
The set $\{\hat{\mathbf{l}}_i \}$ corresponds to the labels predicted by a ML model.
The index $k$ specifies the element of the label vector for which the MAE is computed.
For the purpose of error analysis, we consider $\hat{\mathbf{l}}^{\text{LIF}}$ as the autoencoder's predictions for labels ($\hat{\underline{\mathbf{l}}}_i = \hat{\mathbf{l}}^{\text{LIF}}_i$).
The MAE provides a single number summarizing the overall performance of a model for a given label with index $k$.
However, since different labels have different units, it is not possible to compare the MAE values across different labels.

To compare the performance across different labels, we use the normalized MAE (NMAE).
Compared to MAE, we preceded the computation with a normalization step:
\begin{equation}
\text{NMAE}(\{\mathbf{s}_i, \mathbf{l}_i, \hat{\mathbf{l}}_i \}_{i=1}^{D}, k) = \frac{1}{D} \sum_{i=1}^{D} \frac{\left| \mathbf{l}_i^k - \hat{\mathbf{l}}_i^k \right|}{\sigma_k} ,
\label{eq:NMAE}
\end{equation}
where $\sigma_k$ is the standard deviation of the ground truth labels for the $k$-th label.
This modification enables us to compare performance across different labels, and we utilize it to analyze groups of labels.

\subsection{Reconstruction error}
\label{sec:reconstruction}
Reconstruction error is instrumental in evaluating the model's ability to compress data while preserving key features.
Additionally, this metric helps determine the effectiveness of our models in the ML simulation task.

Measuring reconstruction quality across the entire spectrum, including both the continuum and the spectral lines, presents a challenge due to the relative rarity of line pixels compared to continuum pixels. This difference can lead to misidentifying spectral lines as outliers.

An ideal metric would accurately capture deviations in both the continuum and spectral lines while being robust to outliers.
However, our experiments with traditional data reconstruction metrics reveal a trade-off: Either we fail to fit the continuum adequately (while capturing lines and outliers) or we overfit the continuum and miss both outliers and lines.
Therefore, we chose $\text{MAE}'$ as the balanced metric:
\begin{equation}
\text{MAE}'(\{\mathbf{s}_i, \hat{\mathbf{s}}_i \}_{i=1}^{D}) = \frac{1}{DN} \sum_{i=1}^{D} \sum_{j=1}^{N} \left| \mathbf{s}^j_{i} - \hat{\mathbf{s}}^j_{i} \right|,
\label{eq:MAE_rec}
\end{equation}
where $N$ is the number of pixels in the spectrum, $\mathbf{s}^j_{i}$ is the $j$-th pixel of the $i$-th spectrum, and $\hat{\mathbf{s}}^j_{i}$ is the corresponding prediction.
It is well known that the absolute difference metric in $\text{MAE}'$ is robust to outliers \cite[p.~399]{Murphy_22}.
We experimentally observed that it balances being too robust (insensitive to lines) and not robust enough (sensitive to artifacts).

\subsection{Generative metrics}
\label{sec:generative_metrics}
Generative metrics are an integral part of the evaluation process for models in ML simulation tasks.
The true generative process involves cause-and-effect relationships between the spectral parameters and the resulting spectrum.
However, the learned ML model is not required to replicate these relationships.
It might introduce dependencies between supervised and unsupervised nodes, favor unsupervised nodes for reconstruction, or even ignore supervised nodes.
For example, an ML model could learn to link effective temperature with some of its unsupervised nodes.
Simply adjusting the supervised node that directly represents effective temperature will not be enough for correct simulation.
This is because the model has spread the influence of effective temperature across multiple nodes, complicating how changes in temperature affect the predicted spectrum.

The mapping of labels to spectra and the measurement of reconstruction error cannot identify any of these issues.
Therefore, we need metrics that specifically evaluate the model's ability to simulate spectra based on interventions in individual labels, thus accurately measuring the cause-and-effect relationships between isolated spectral parameters and the resulting spectrum.

Our chosen approach is inspired by the latent space traversal \citep{infoGAN},
which is a recognized method for qualitative evaluation of generative properties.
In latent space traversal, we manually modify the latent representation and observe the effects on the output spectrum.
For example, we can alter the node responsible for the radial velocity and observe a Doppler shift in the output spectrum, provided the node captures the essence of radial velocity.
This approach provides insight into the connection between labels and spectra.
However, the challenge lies in quantifying this intuition.

Quantification of latent space traversal can be achieved by automatically intervening in the spectral parameters and measuring the reconstruction error against the true generative process, denoted as $M$.
To facilitate this, we introduce a $\doop$ operator to represent interventions on specific spectral parameters:
\begin{equation}
    \mathbf{b}_i' = \doop(\mathbf{b}_i, \triangle l^k_{\text{scaled}}, k),
\end{equation}
where $\mathbf{b}_i$ denotes the latent representation of the $i$-th spectrum by some encoder $q$, $k$ refers to the index of the node that is associated with the intervened spectral parameter, and $\triangle l^k_{\text{scaled}}$ is the intervention value for the spectral parameter (scaled using Eq.~\eqref{eq:norm}).
The operation $\doop(\mathbf{b}_i, \triangle l^k_{\text{scaled}}, k)$ modifies the $k$-th element of $\mathbf{b}_i$ by adding the value $\triangle l^k_{\text{scaled}}$, resulting in the altered latent representation $\mathbf{b}_i'$.    

Our first generative metric is specific to radial velocity since we can model $M$ by the operator $\shift(\mathbf{s}, v)$ that applies a Doppler shift to the spectrum $\mathbf{s}$ based on the radial velocity $v$.
This approach is valid for real data if we carefully select a range of wavelengths that are minimally affected by telluric lines, which are not affected by shift in a star's radial velocity.
We defined the radial velocity intervention score (RVIS) metric as the reconstruction error between Doppler shifting a spectrum $\mathbf{s}$ and using the node associated with  radial velocity to shift the same spectrum $\mathbf{s}$:
\begin{equation}
    \text{RVIS} = \mathbb{E}_{i, \triangle v \in V, \lambda \in \Lambda}[|p \circ \doop(\mathbf{b}_i, \triangle v_{\text{scaled}}, k_v)(\lambda) - \shift(\hat{\mathbf{s}}_{i}, \triangle v)(\lambda)|].
    \label{eq:rvis},
\end{equation}
\sloppy
where $i$ is the index of the spectrum in the dataset, $\triangle v$ is the radial velocity intervention, $V \in [-40, 40] \, \si{\kilo\meter\per\second}$ is a uniformly sampled set of considered shifts, $\lambda$ is a wavelength, $\Lambda \in [6050, 6250] \, \si{\angstrom}$ is a narrow wavelength range that is minimally affected by the telluric lines, $k_v$ is the index of the node associated with radial velocity, $p$ is any decoder capable of projecting the vector $\mathbf{b}$ onto a spectrum $\mathbf{s}$,
the operator $\circ$ represents the composition of functions, and $\hat{\mathbf{s}}_{i}$ represents the output without any intervention ($\triangle v = 0$).
This metric is evaluated on the HARPS test dataset.

\fussy
For the other labels, we have to utilize the simulation data and the known associated $M$.
Specifically, we use ETC simulations to evaluate the generative properties of ML models for temperature, metallicity, and surface gravity.
The generative metric is called the general intervention score (GIS) and is defined as follows:
\begin{equation}
    \text{GIS}(k) = \mathbb{E}_{i, \triangle l^k, \lambda}\bigl[|p \circ \doop(\mathbf{b}_i, \triangle l^k_{\text{scaled}}, k)(\lambda) - p \circ q \circ M \circ \doop(\mathbf{l}, \triangle l^k, k) (\lambda)|\bigr].
    \label{eq:gis}.
\end{equation}
The first term in Eq.~\eqref{eq:gis} gives us reconstruction of a spectrum, when constraining modification to the target node-label pair.
The second term gives us reconstruction, when the ML model is unconstrained.
Hence, the score indicates how the average reconstruction error suffer when cause-and-effect relationship is enforced during decoding.
If $q$ is a classic encoder, we simply send the output to the decoder $p$. 
Alternatively, if $q$ is a probabilistic encoder, we first randomly sample the latent representation from it.

In this study, we evaluate the GIS using the generative ETC dataset described in Sect.~\ref{sec:data}, wherein a single label changes while the remaining ones stay constant.
Hence, we can automatize traversing individual labels.

\subsection{Model selection}
\label{sec:bottleneck_optimization}
Model selection in machine learning refers to the process of evaluating a specific metric on a validation dataset for a set of model candidates and selecting the best model based on the chosen metric.
The candidates may include various settings of ML hyperparameters, such as modifications to the model's architecture or training process, as well as different types of models.
The metric used is often the same loss function that is utilized during training.
In our case, we would like metrics that can relate our models back to our inference tasks described in Sect.~\ref{sec:inference}, specifically, label prediction and ML simulation.
This is quite straightforward for label prediction, where we simply use the supervised term introduced in Eq.~\eqref{eq:lab}.

Model selection for ML simulation is far more challenging since the reconstruction error is not suitable for model selection.
The critical ML hyperparameter for an AE is the bottleneck size; however, we cannot use the reconstruction error as a metric because it generally decreases as the bottleneck size increases.
Constants $\lambda_\mathcal{KL}$ in $\beta$-VAE, and $\lambda_\text{MI}$ and $\lambda_\text{MMD}$ in infoVAE loss functions also serve as regularization of the bottleneck, similar to using different bottleneck sizes in AEs.
Therefore, we minimize reconstruction error by setting all constants to zero, which results in a collapse of the advanced models into the AE model.
Without a suitable metric, the model selection process requires careful manual analysis of each model candidate \citep{Nima_2021}.
This process is computationally expensive and time-consuming, especially when the dataset is large, the model is complex, and there is a need to compare many candidates.

Therefore, we chose the RVIS metric in Eq.\eqref{eq:rvis} as the model selection metric because it provides a direct interpretation and exhibits a convex curve with respect to ML hyperparameters, allowing for the selection of an optimum without a computationally costly grid search.
We expect similar behavior for the GIS metric in Eq.~\eqref{eq:gis}, but we have not investigated this, as we cannot evaluate it on real data.

\begin{figure}[htbp]
    \centering
    \includegraphics[width=0.5\textwidth]{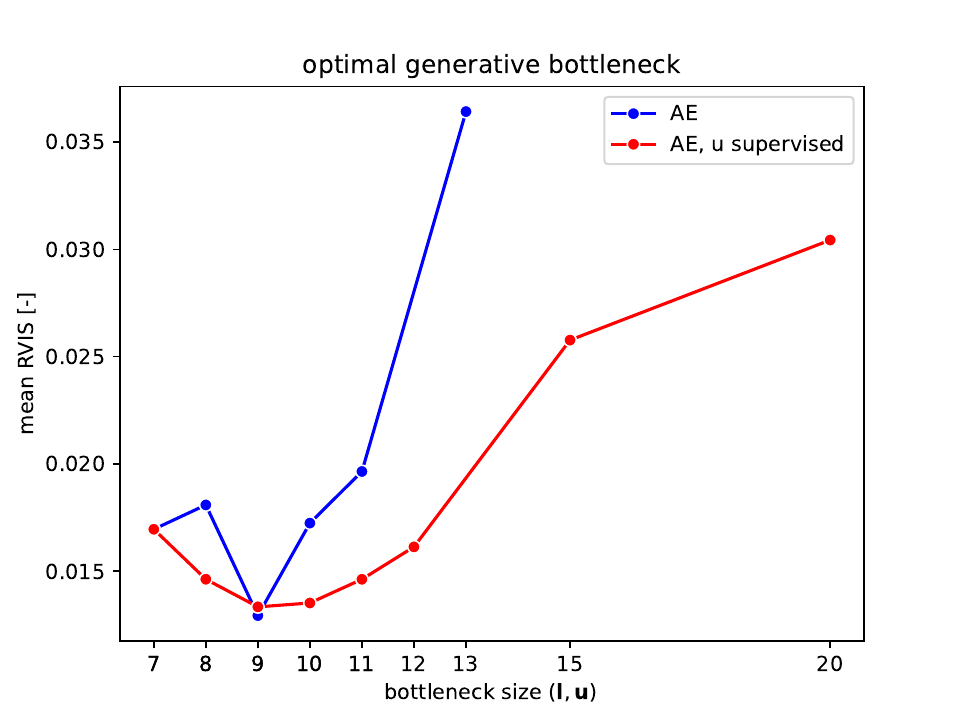}
    \caption{
        Relation between bottleneck size and RVIS.
        The vector $\mathbf{l}$ size is fixed to seven.
        In addition, vector $\mathbf{u}$, which represents the unsupervised portion of the latent representation, varies in size from zero to 13.
    }
    \label{fig:AE_bottleneck_RVIS}
\end{figure}

Figure~\ref{fig:AE_bottleneck_RVIS} presents our experimentation with various sizes of $\mathbf{u}$ for AEs.
The size of $\mathbf{l}$ is fixed to seven. 
All models were trained on a mixture of real and ETC data.
In the graph, each point represents a single experiment; the x-axis shows the total size of the bottleneck, while the y-axis shows the RVIS metric from Eq.~\eqref{eq:rvis}, evaluated using a testing real dataset.
The red curve illustrates results from experiments in which we supervise $\mathbf{u} = 0$ for ETC data, as described in Sect.\ref{sec:augmentation}.
Conversely, the blue curve represents results from experiments conducted without this specific augmentation.

We interpret the results in Fig.~\ref{fig:AE_bottleneck_RVIS} as follows:
A bottleneck size of nine appears to be optimal.
For the bottleneck with just seven or eight nodes, the node responsible for radial velocity is pushed to contain additional information, which leads to the worsening of the RVIS performance.
When the bottleneck is larger, the dispersion of radial velocity information between multiple nodes also results in an increase in RVIS.

While the overall trends are not altered by the augmentation ($\mathbf{u} = 0$), the red curve is noticeably flatter.
Hence, the suggested augmentation alleviates the RVIS sensitivity to the size of the bottleneck.
As a result, it allows us to slightly increase the bottleneck size without significant penalization.
As shown in Fig.~\ref{fig:AE_bottleneck_RVIS}, increasing the bottleneck size to 10 minimally impacts RVIS.

We set the hyperparameter $\lambda_\text{lab}$, which balances reconstruction loss and label loss, to 1 based on initial experiments showing that label prediction performance is relatively insensitive to its value.
Results for other values of $\lambda_\text{lab}$ and their impact on reconstruction and label prediction are presented in Sect.~\ref{sec:lambda_lab}.

In order to configure the ML hyperparameters for the supervised $\beta$-VAE models, we proceeded in the following way.
First, we set the bottleneck to a fixed size of 32 nodes, of which seven were supervised and the remaining 25 were not.
We searched for a $\lambda_{\mathbb{KL}}$ that minimizes the RVIS metric, eventually selecting $\lambda_{\mathbb{KL}} = 0.00001$.
The same value of $\lambda_{\mathbb{KL}}$ was used for the $\beta$-VAE (U) model, in which all nodes are unsupervised.

For the supervised infoVAE model, we initially set $\lambda_{\text{MI}}$ to $1 - \lambda_{\mathbb{KL}}$ and $\lambda_{\text{MMD}}$ to $1 - \lambda_{\text{MI}}$, essentially using the $\beta$-VAE model, given that the third term in Eq.~\eqref{eq:infoVAE_equivalent} is nullified by these ML hyperparameter settings.
The bottleneck configuration was identical to that of the supervised $\beta$-VAE model.
Subsequently, we increased $\lambda_{\text{MMD}}$ until the RVIS metric was minimized, establishing that $\lambda_{\text{MMD}}$ should be set to $100$.
The same values of $\lambda_{\mathbb{MI}}$ and $\lambda_{\mathbb{MMD}}$ were applied to the unsupervised infoVAE (U) model.

\section{Results}
\label{sec:results}
In this section, we present the results of our experiments.
We investigate our two main objectives: label prediction and ML simulation, as detailed in Sect.~\ref{sec:inference}.
The evaluation of label prediction is straightforward, as we have access to the catalog that we consider the source of ground truth.
We focus on the performance under varying conditions to better understand the contribution of individual components.

The summary of results for label prediction is in Table~\ref{tab:summary}.
The summary table contains a selection showcasing the most significant results with practical implications.
It also includes references to other tables that present additional nuances.

Conversely, the evaluation of ML simulation remains challenging due to the semi-supervised nature of the training.
The full extent of the challenges associated with ML simulation evaluation is explained in Sect.~\ref{sec:ML_simulation}.
Here, we focus on establishing an appropriate evaluation approach for ML simulations.

In the remaining sections, we relate the results to applications in spectroscopy.
Therefore, we occasionally refrain from using the ML term labels that were defined in Sect.~\ref{sec:intro}.
Instead, we specify that we discuss stellar, spectral, or physical parameters.

\begin{table*}[ht]
    \center
    \caption{
    Summary comparison of supervised, semi-supervised, and unsupervised models in this work for label prediction.
    }
    \label{tab:summary}
    \small
    \begin{tabular}{llllllll}
    \toprule
    & \multirow{2}{*}{\shortstack[l]{$T_{\rm eff}$\\$(\mathrm{K})$}} & \multirow{2}{*}{\shortstack[l]{[M/H]\\($\mathrm{dex}$)}} & \multirow{2}{*}{\shortstack[l]{$\log(g)$\\($\mathrm{dex}$)}} & Available labels & Dataset & Details\tablefootmark{a} \\
     Models  & & & & (real data) & & \\
     \midrule
    supervised and semi-supervised & $\sim50$ & $\sim0.03$ & $\sim0.04$ & 100\% & mixed & Table~\ref{tab:results_main}  \\ 
    supervised & $\sim155$ & $\sim0.05$ & $\sim0.07$ & 1\% & real &  Table~\ref{tab:results_auxiliary} \\ 
    semi-supervised & $\sim75$ & $\sim0.03$ & $\sim0.06$ & 1\% & mixed & Table~\ref{tab:results_auxiliary} \\ 
    unsupervised & $\sim1000$ & $\sim0.17$ & $\sim0.18$ & 0\% & real & Table~\ref{tab:results_unsupervised} \\ 
    \bottomrule
\end{tabular}
\label{tab:model_performance}
\tablefoot{
\tablefoottext{a}{The referenced tables provide the source models.}
}
\end{table*}

\subsection{Label prediction}
\label{sec:label_prediction}
\sloppy
An important application of our ML models is to predict physical parameters of the stars from the data.
For that purpose, we tested various models to predict labels: stand-alone encoders from Sect.~\ref{sec:supervised}, AEs from Sect.~\ref{sec:AE} and Sect.~\ref{sec:semi_supervised}, and VAEs from Sect.~\ref{sec:infoVAE}.
Models were trained for 1000 epochs using Adam Optimizer with a learning rate of $10^{-4}$.
Both AEs and VAEs used a decoder pretrained on ETC data.

\fussy
We evaluated all models on the HARPS test dataset described above, categorizing label errors into intrinsic to the star (temperature, metallicity, and surface gravity) and extrinsic (radial velocity, BERV, and airmass) groups. 
Individual label errors were calculated using Mean Absolute Error (MAE, Eq.~\eqref{eq:MAE}), and overall errors ("All") for each label group were determined using Normalized Mean Absolute Error (NMAE, Eq.~\eqref{eq:NMAE}).

Table~\ref{tab:results_main} presents various prediction errors across different models, including those for temperature and metallicity.
The unsupervised model $\beta$-VAE (U*) from \cite{Nima_2021}, which relies on a posteriori training for label prediction, produced a temperature error of $978.0 \pm 13.0$~K and a metallicity error of $0.1578 \pm 0.0027$~dex, using essentially the same data employed in this work. 
In contrast, the supervised model CNN (mix) achieved a temperature error of only $50.21 \pm 0.82$~K and a metallicity error of $0.02118 \pm 0.00064$~dex, demonstrating improvements by nearly a factor of $20$ and $7.5$, respectively, over the unsupervised model.
These findings underscore that label-aware (supervised or semi-supervised) representation learning is crucial for accurate label prediction, achieving significant improvements on par with the best traditional methods \citep[e.g.,][]{miller_2020}.

\begin{figure*}
    \centering
    \begin{subfigure}{0.33\textwidth} \includegraphics[width=\textwidth]{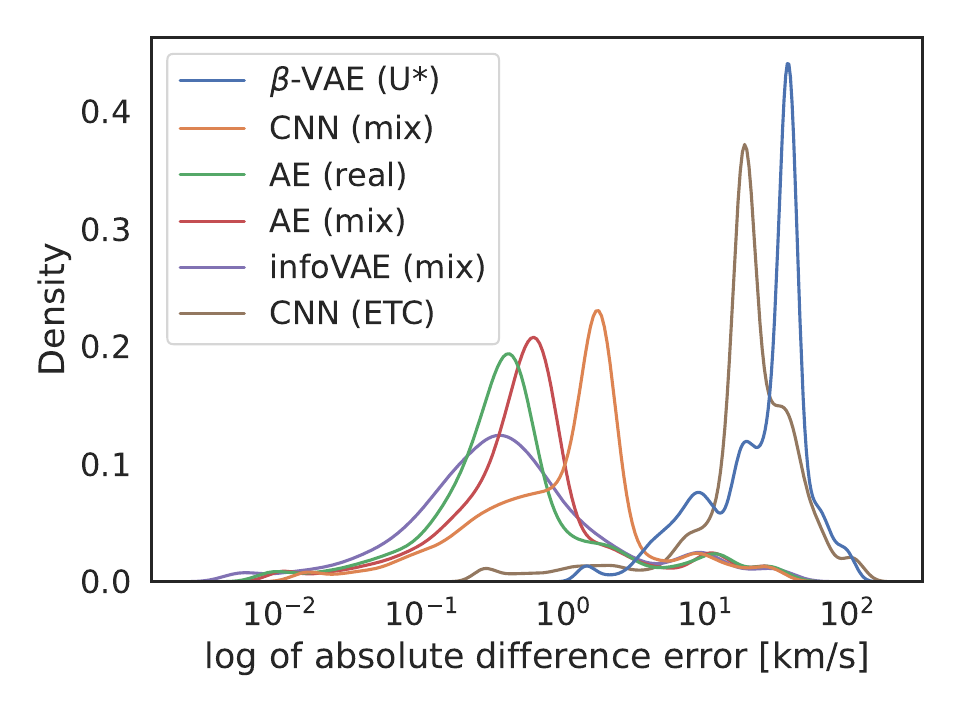}
        \caption{Radial velocity [\si{\kilo\meter\per\second}]}
    \end{subfigure}
    \begin{subfigure}{0.33\textwidth}
        \includegraphics[width=\textwidth]{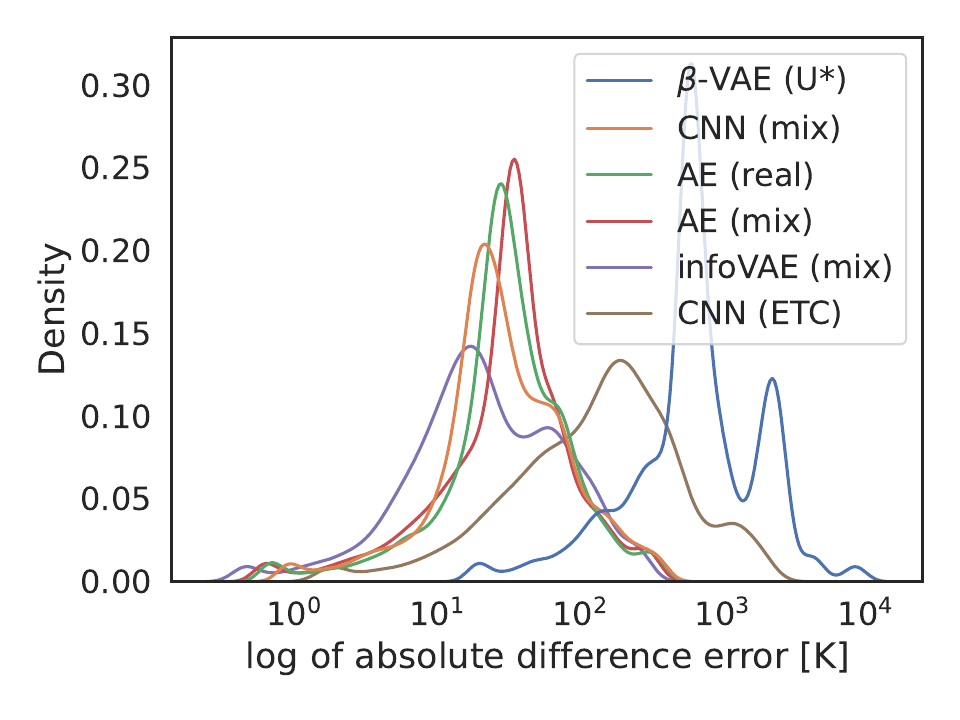}
        \caption{Effective Temperature ($T_\text{eff}$) [K]}
    \end{subfigure}
    \begin{subfigure}{0.33\textwidth} \includegraphics[width=\textwidth]{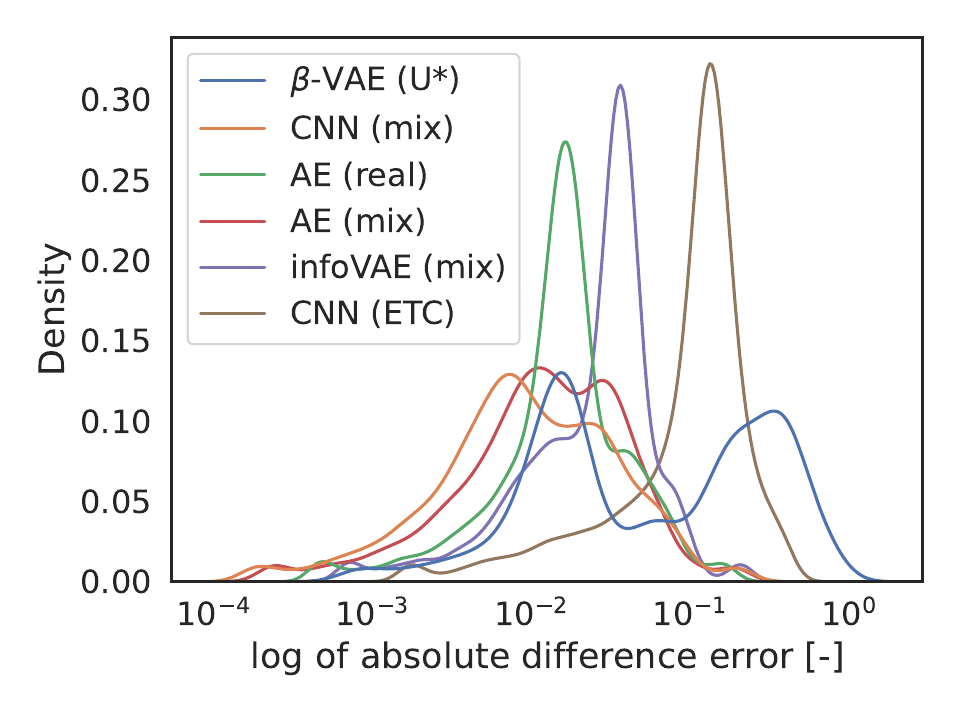}
        \caption{Metallicity ([M/H])}
    \end{subfigure}
    \begin{subfigure}{0.33\textwidth}
        \includegraphics[width=\textwidth]{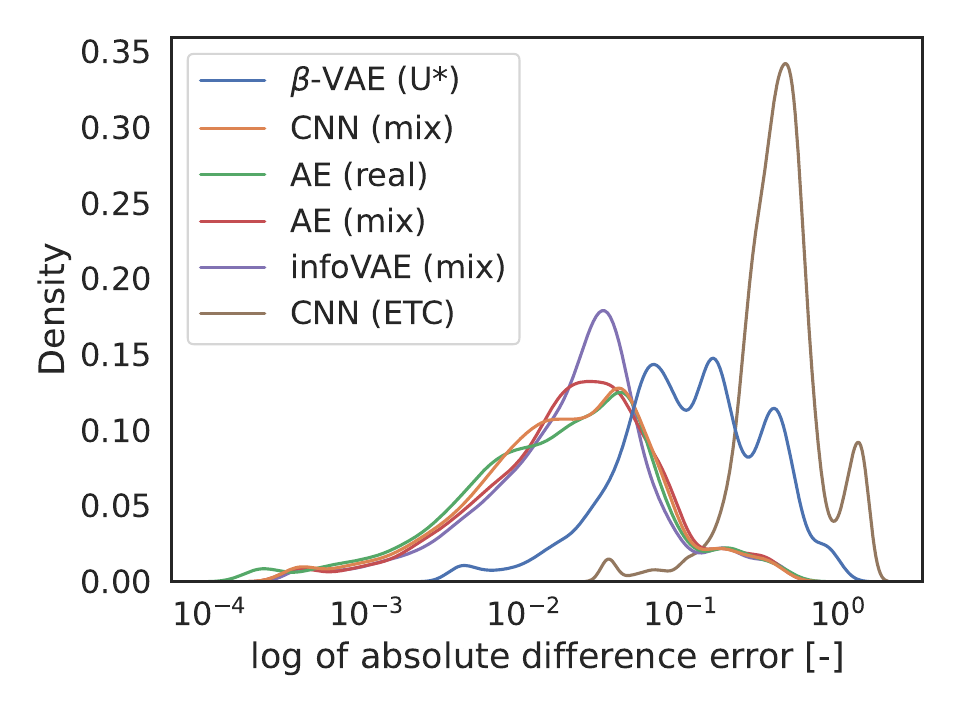}
        \caption{Gravity ($\log g$)}
    \end{subfigure}
    \begin{subfigure}{0.33\textwidth}
        \includegraphics[width=\textwidth]{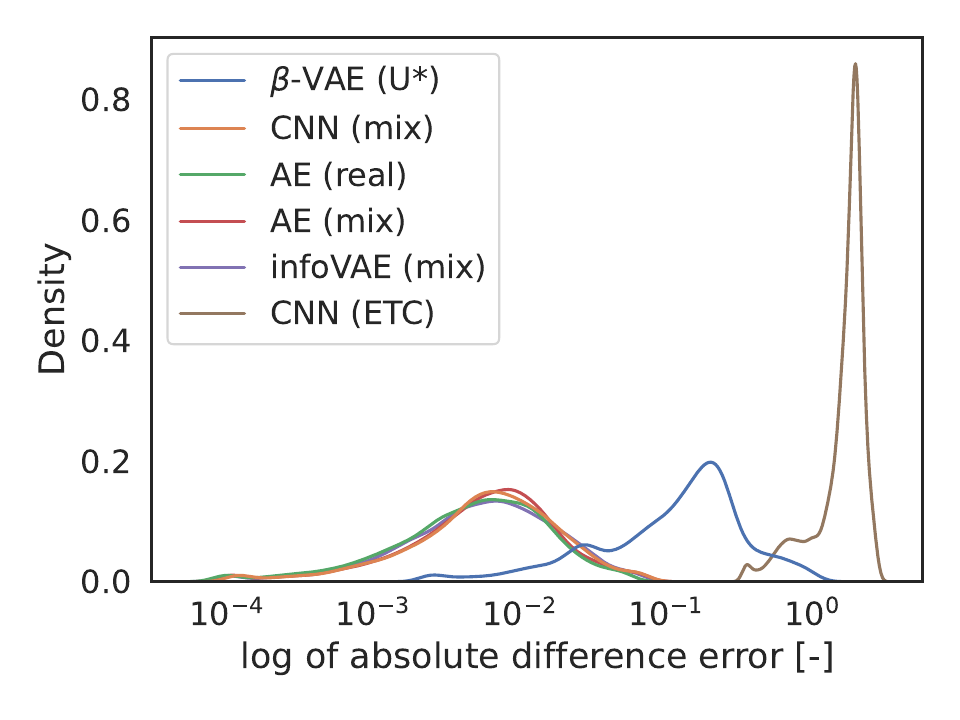}
        \caption{Airmass}
    \end{subfigure}
    \caption{
    Kernel density estimation plots illustrating the distribution of absolute error differences.
    The KDE bandwidth is determined by Scott's rule and is clipped between the first and 99th percentiles.
    The models are supervised encoders (real and mixed data), supervised AE (bottleneck = 9), supervised infoVAE (bottleneck=32), and VAE (bottleneck = 128) \citep{Nima_2021}.
    }
    \label{fig:labels_detail}
\end{figure*}

Each physical parameter achieves the best prediction with a different model in Table~\ref{tab:results_main}, which could be a consequence of degeneracy, where different combinations of physical parameters lead to similar spectra.
This suspicion is partially supported by MAE distributions in Fig.~\ref{fig:labels_detail}.
Most prominently, infoVAE (mix) appears multimodal for temperature (where it also achieves the best performance with an MAE of $43.68 \pm 0.73$ K).
The metallicity error distributions also show multimodality, especially for AE (mix) and CNN (mix) (the best model with an MAE of $0.02118 \pm 0.00064$ dex).
Radial velocity and airmass do not demonstrate similar behavior.
Further experimentation is necessary to support the degeneracy claim, as alternative explanations exist for the observed multimodalities, such as clusters in the data.

\begin{figure}
    \centering
    \begin{subfigure}[b]{0.49\textwidth}
        \includegraphics[width=\textwidth]{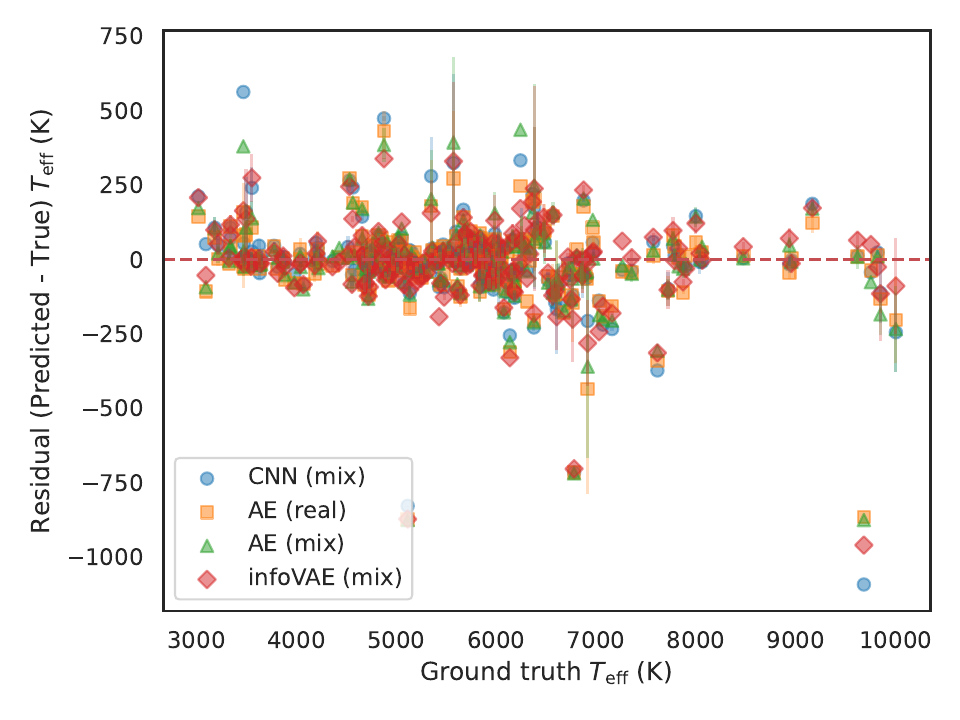}
        \caption{$T_\text{eff}$ [K]}
    \end{subfigure}
    \begin{subfigure}[b]{0.49\textwidth} \includegraphics[width=\textwidth]{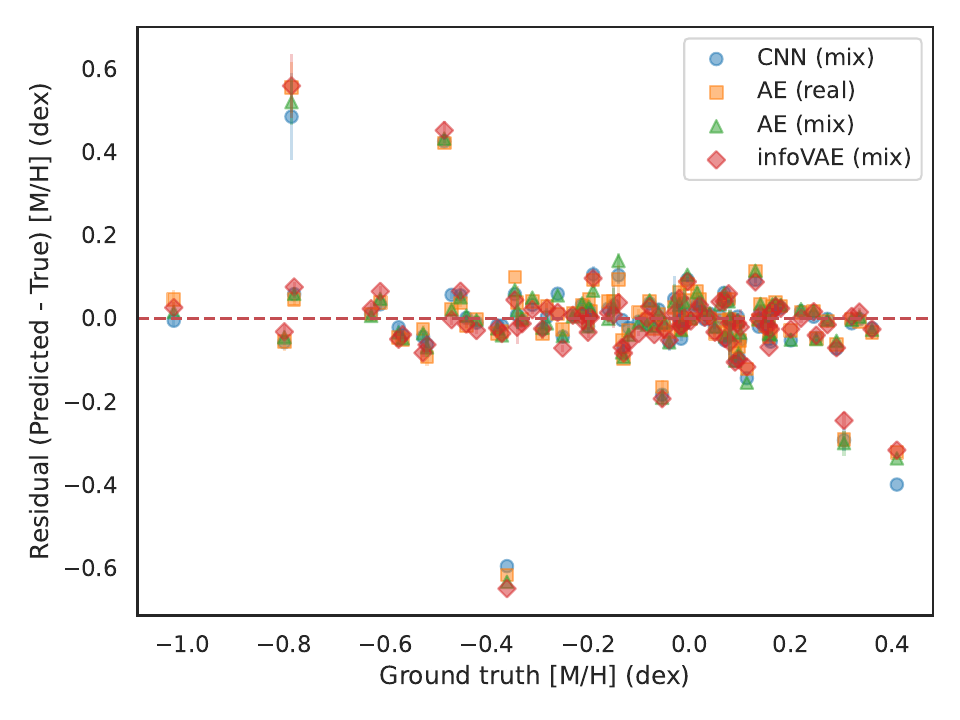}
        \caption{[M/H]}
    \end{subfigure}
    \begin{subfigure}[b]{0.49\textwidth}
        \includegraphics[width=\textwidth]{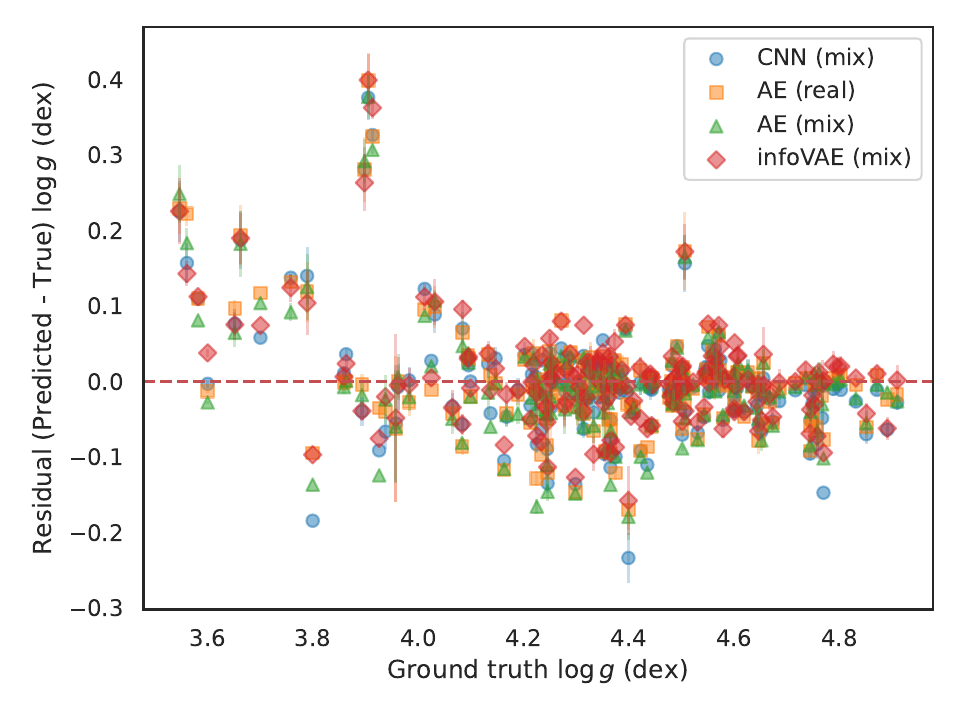}
        \caption{$\log g$}
    \end{subfigure}
    \caption{
        Residual plots for effective temperature ($T_\text{eff}$), metallicity ([M/H]), and surface gravity ($\log g$) predictions across different models.
    }
    \label{fig:residuals}
\end{figure}

The residual plots in Fig.~\ref{fig:residuals} illustrate the relationship between the true and predicted labels as a function of the true label for selected models from Table~\ref{tab:results_main}: CNN (mix), AE (real), AE (mix), and infoVAE (mix).
The x-axis shows the ground truth values, while the y-axis displays the residuals (predicted minus true values).
The plots show that model performance decreases in sparsely populated regions.
In particular, we observe a decline in performance for high temperatures (above 10,000~K) and low surface gravity (below 3.8~dex).
These observations are consistent with the bias observed in Fig.~\ref{fig:teff_logg_plot}.
Additionally, there is no significant difference between the various models in terms of residuals.
Since these parameters are also challenging for classical methods, the reliability of the catalog values is questionable, further complicating the analysis.

Table~\ref{tab:results_main} indicates that the CNN (ETC) outperforms the $\beta$-VAE (U*), specifically in the cases of temperature, radial velocity, and metallicity.
We suspect that this superiority is due to the critical roles of temperature (a prime factor influencing the continuum) and radial velocity (which shifts every pixel in the spectrum) in achieving accurate reconstruction.
However, the CNN (ETC) model performs significantly worse than all the other label-aware models, as shown in Tables~\ref{tab:results_main} and \ref{tab:results_auxiliary}.
This observation suggests that models trained solely on simulated data may be inadequately equipped to handle the variations encountered in real-world data.

\begin{table*}[ht]
    \caption{
    Overview of learning paradigms and architectures and their corresponding MAE for various labels.
    }
    \label{tab:results_main}
    \centering
    \begin{subtable}{\textwidth}
        \centering
        \subcaption{Model Descriptions}
        \small
        \begin{tabular}{llllrl}
\toprule
 & Model type & Learning paradigm & Architecture & Bottleneck & Dataset \\
Legend descriptor &  &  &  &  &  \\
\midrule
$\beta$-VAE (U*) (1) & $\beta$-VAE & unsupervised & CNN-CNN & 128 & real \\
CNN (mix) & encoder & supervised & CNN & 6 & mixed \\
AE (real) & AE & semi-supervised & CNN-ResNet & 9 & real \\
AE (mix) & AE & semi-supervised & CNN-ResNet & 9 & mixed \\
infoVAE (mix) & infoVAE & semi-supervised & CNN-ResNet & 32 & mixed \\
CNN (ETC) & encoder & supervised & CNN & 7 & ETC \\
\bottomrule
\end{tabular}

    \end{subtable}
    \hfill
    \vspace{1mm}
    \begin{subtable}{\textwidth}
        \centering
        \subcaption{Intrinsic Labels}
        \small
        \begin{tabular}{lllll}
\toprule

 & \multirow{2}{*}{\shortstack[l]{$T_{\rm eff}$\\$(\mathrm{K})$}} & \multirow{2}{*}{\shortstack[l]{[M/H]\\($\mathrm{dex}$)}} & \multirow{2}{*}{\shortstack[l]{$\log(g)$\\($\mathrm{dex}$)}} & \multirow{2}{*}{\shortstack[l]{All\\(-)}} \\
 & & & & \\
\midrule
$\beta$-VAE (U*) (1) & 978.0 ± 13.0 & 0.1578 ± 0.0027 & 0.1798 ± 0.0025 & 1.548 ± 0.013 \\
CNN (mix) & 50.21 ± 0.82 & \textbf{0.02118 ± 0.00064} & 0.03957 ± 0.00085 & \textbf{0.1793 ± 0.0022} \\
AE (real) & 50.32 ± 0.83 & 0.02478 ± 0.00056 & \textbf{0.03898 ± 0.00089} & 0.1844 ± 0.0022 \\
AE (mix) & 49.74 ± 0.77 & 0.02292 ± 0.00055 & 0.04139 ± 0.00087 & 0.1867 ± 0.0021 \\
infoVAE (mix) & \textbf{43.68 ± 0.73} & 0.03424 ± 0.00058 & 0.04005 ± 0.00087 & 0.1986 ± 0.0021 \\
CNN (ETC) & 273.1 ± 4.2 & 0.1222 ± 0.0011 & 0.4994 ± 0.0044 & 1.7029 ± 0.0091 \\
\bottomrule
\end{tabular}

    \end{subtable}
    \hfill
    \vspace{1mm}
    \begin{subtable}{\textwidth}
        \centering
        \subcaption{Extrinsic Labels}
        \small
        \begin{tabular}{lllll}
\toprule
 &  \multirow{2}{*}{\shortstack[l]{Radvel\\(\si{\kilo\meter\per\second})}}  & \multirow{2}{*}{\shortstack[l]{BERV\\(\si{\kilo\meter\per\second})}}  & \multirow{2}{*}{\shortstack[l]{Airmass\\(-)}} & \multirow{2}{*}{\shortstack[l]{All\\(-)}} \\
 & & & & \\
\midrule
$\beta$-VAE (U*) (1) & 31.07 ± 0.27 & 16.01 ± 0.096 & 0.181 ± 0.0019 & 1.5751 ± 0.0074 \\
CNN (mix) & 2.375 ± 0.061 & 0.1671 ± 0.002 & 0.01069 ± 0.00022 & 0.0815 ± 0.001 \\
AE (real) & 1.94 ± 0.069 & \textbf{0.1214 ± 0.0021} & \textbf{0.00885 ± 0.00018} & \textbf{0.0664 ± 0.0011} \\
AE (mix) & \textbf{1.918 ± 0.063} & 0.1697 ± 0.0024 & 0.0103 ± 0.00022 & 0.0709 ± 0.0011 \\
infoVAE (mix) & 1.967 ± 0.071 & 0.1855 ± 0.0024 & 0.01023 ± 0.00021 & 0.0722 ± 0.0011 \\
CNN (ETC) & 25.43 ± 0.27 & NaN ± NaN & 1.6827 ± 0.0054 & 6.932 ± 0.016 \\
\bottomrule
\end{tabular}
\end{subtable}
\tablefoot{
The column titled "All" uses NMAE from Eq.~\eqref{eq:NMAE} to summarize all labels in the group.
The notation "a ± b" represents the mean absolute error ± the standard deviation of the estimate for each respective label and model.
Using boldface signifies that the model outperformed the other models significantly for the label in that column. This significance was determined with a significance level of $\alpha = 0.01$ using Mann-Whitney U tests with Holm–Bonferroni correction for multiple comparisons \citep{Mann_Whitney_U_test, Holm_Bonferroni_correction}.
}
\tablebib{(1)~\citet{Nima_2021}.}
\end{table*}

Table~\ref{tab:results_auxiliary} examines how reducing the quantity of the label and considering different datasets and model types affect the mean absolute label error.
The "legend descriptor" column specifies the model type (CNN or AE), the percentage of catalog labels utilized in training (100\% or 1\%), and the dataset employed (real or mixed).

The results from employing the full dataset of real labels are detailed in Table~\ref{tab:results_auxiliary}.
The CNN 100\% (real) model achieves an intrinsic error of $0.1837 \pm 0.0022$, while the CNN 100\% (mix) model slightly improves with an intrinsic error of $0.1793 \pm 0.0022$.
Conversely, the extrinsic error increases in the mixed model, with the CNN 100\% (real) model at $0.07591 \pm 0.00098$ compared to $0.0815 \pm 0.001$ for the CNN 100\% (mix) model.
Semi-supervision yields mixed results; temperature labels improve, but radial velocity labels exhibit a negative trend.

These findings indicate that potential benefits from mixed data or semi-supervision are small and inconclusive if we have a large dataset with labels.
The efficacy of semi-supervision varies depending on the category of the label being predicted, suggesting that while mixed data approaches do not degrade network performance, their advantages are inconsistent across all label types and error metrics.
This implies that semi-supervised learning or simulated data are unnecessary for label prediction if we have sufficient data.

Table~\ref{tab:results_auxiliary} shows that the model CNN 1\% (mix) achieves a temperature prediction error of $108.6 \pm 5.6$ when trained with just 1\% of available labels.
In comparison, the model trained exclusively on real data has a higher error of $154.4 \pm 5$.
This pattern is consistent across different model types and label groups, as evidenced by the CNN 1\% (mix) model's intrinsic error of $0.2984 \pm 0.0051$, which is superior to the CNN 1\% (real) model's error of $0.3954 \pm 0.0061$.
Moreover, even with only 1\% of the available labels, effective temperature reconstruction remains accurate enough for various astrophysical applications.
These findings expose the role of simulated data in boosting label prediction accuracy in scenarios with limited real data, while maintaining equivalent efficacy in cases where real data are abundant.

Table~\ref{tab:results_auxiliary} also compares supervised CNN 1\% (mix) and semi-supervised AE 1\% (mix).
The most striking difference is the prediction of BERV, which is $0.787 \pm 0.011$ for CNN 1\% (mix) and $0.3719 \pm 0.0062$ for AE 1\% (mix).
However, there is also a significant improvement for temperatures ($108.6 \pm 5.6$ vs. $75.2 \pm 2.8$), metallicity ($0.04468 \pm 0.00081$ vs. $0.02934 \pm 0.00081$), radial velocity ($3.91 \pm 0.14$ vs. $3.53 \pm 0.15$), and overall intrinsic labels ($0.2984 \pm 0.0051$ vs. $0.2578 \pm 0.0037$).
The only reason why CNN 1\% (mix) performs better on overall extrinsic labels is that airmass dominates this metric due to the normalization.
Our results suggest that semi-supervision is another source of improvement when real data are scarce and we have access to simulated data.

\begin{table*}
    \caption{
        Effects of reduced label quantity and mixed datasets on mean absolute errors for various labels.
    }
    \label{tab:results_auxiliary}
    \centering
    \begin{subtable}{\textwidth}
        \centering
        \subcaption{Model Descriptions}
        \small
        \begin{tabular}{lllllr}
\toprule
 & Model type & Learning paradigm & Architecture & Dataset & \multirow{2}{*}{\shortstack[l]{Available labels\\(real data)}} \\
Legend descriptor &  &  &  &  &  \\
\midrule
CNN 1.00 (real) & encoder & supervised & CNN & real & 100\% \\
CNN 0.01 (real) & encoder & supervised & CNN & real & 1\% \\
CNN 1.00 (mix) & encoder & supervised & CNN & mixed & 100\% \\
CNN 0.01 (mix) & encoder & supervised & CNN & mixed & 1\% \\
AE 1.00 (real) & AE & semi-supervised & CNN-ResNet & real & 100\% \\
AE 0.01 (real) & AE & semi-supervised & CNN-ResNet & real & 1\% \\
AE 1.00 (mix) & AE & semi-supervised & CNN-ResNet & mixed & 100\% \\
AE 0.01 (mix) & AE & semi-supervised & CNN-ResNet & mixed & 1\% \\
\bottomrule
\end{tabular}

    \end{subtable}
    \hfill
    \vspace{1mm}
    \begin{subtable}{\textwidth}
        \centering
        \subcaption{Intrinsic Labels}
        \small
        \begin{tabular}{lllll}
\toprule
 & \multirow{2}{*}{\shortstack[l]{$T_{\rm eff}$\\$(\mathrm{K})$}} & \multirow{2}{*}{\shortstack[l]{[M/H]\\($\mathrm{dex}$)}} & \multirow{2}{*}{\shortstack[l]{$\log(g)$\\($\mathrm{dex}$)}} & \multirow{2}{*}{\shortstack[l]{All\\(-)}} \\
 & & & & \\
\midrule
CNN 1.00 (real) & 51.27 ± 0.81 & 0.02109 ± 0.00064 & 0.04102 ± 0.00085 & 0.1837 ± 0.0022 \\
CNN 0.01 (real) & 154.4 ± 5.8 & 0.049 ± 0.0012 & 0.0712 ± 0.0014 & 0.3954 ± 0.0061 \\
CNN 1.00 (mix) & 50.21 ± 0.82 & 0.02118 ± 0.00064 & 0.03957 ± 0.00085 & 0.1793 ± 0.0022 \\
CNN 0.01 (mix) & 108.6 ± 5.6 & 0.04468 ± 0.00081 & 0.05083 ± 0.00098 & 0.2984 ± 0.0051 \\
AE 1.00 (real) & 50.32 ± 0.83 & 0.02478 ± 0.00056 & 0.03898 ± 0.00089 & 0.1844 ± 0.0022 \\
AE 0.01 (real) & 100.2 ± 1.7 & 0.082 ± 0.0012 & 0.0848 ± 0.0013 & 0.4439 ± 0.0038 \\
AE 1.00 (mix) & 49.74 ± 0.77 & 0.02292 ± 0.00055 & 0.04139 ± 0.00087 & 0.1867 ± 0.0021 \\
AE 0.01 (mix) & 75.2 ± 2.8 & 0.02934 ± 0.00081 & 0.0567 ± 0.0011 & 0.2578 ± 0.0037 \\
\bottomrule
\end{tabular}

    \end{subtable}
    \hfill
    \vspace{1mm}
    \begin{subtable}{\textwidth}
        \centering
        \subcaption{Extrinsic Labels}
        \small
        \begin{tabular}{lllll}
\toprule
 &  \multirow{2}{*}{\shortstack[l]{Radvel\\(\si{\kilo\meter\per\second})}}  & \multirow{2}{*}{\shortstack[l]{BERV\\(\si{\kilo\meter\per\second})}}  & \multirow{2}{*}{\shortstack[l]{Airmass\\(-)}} & \multirow{2}{*}{\shortstack[l]{All\\(-)}} \\
 & & & & \\
\midrule
CNN 1.00 (real) & 1.994 ± 0.058 & 0.1764 ± 0.0018 & 0.01163 ± 0.00022 & 0.07591 ± 0.00098 \\
CNN 0.01 (real) & 4.4 ± 0.19 & 0.7122 ± 0.0071 & 0.06974 ± 0.00066 & 0.2839 ± 0.0031 \\
CNN 1.00 (mix) & 2.375 ± 0.061 & 0.1671 ± 0.002 & 0.01069 ± 0.00022 & 0.0815 ± 0.001 \\
CNN 0.01 (mix) & 3.91 ± 0.14 & 0.787 ± 0.011 & 0.04042 ± 0.00063 & 0.2047 ± 0.0026 \\
AE 1.00 (real) & 1.94 ± 0.069 & 0.1214 ± 0.0021 & 0.00885 ± 0.00018 & 0.0664 ± 0.0011 \\
AE 0.01 (real) & 5.19 ± 0.19 & 0.4659 ± 0.0063 & 0.07247 ± 0.00073 & 0.3002 ± 0.0032 \\
AE 1.00 (mix) & 1.918 ± 0.063 & 0.1697 ± 0.0024 & 0.0103 ± 0.00022 & 0.0709 ± 0.0011 \\
AE 0.01 (mix) & 3.53 ± 0.15 & 0.3719 ± 0.0062 & 0.05555 ± 0.00058 & 0.221 ± 0.0026 \\
\bottomrule
\end{tabular}
\end{subtable}
\tablefoot{
The first table provides an overview of the model properties.
The column titled "All" uses NMAE to summarize the mean absolute error across all normalized labels.
The column named "Available labels (real data)" indicates the percentage of HARPS labels in the catalog that were used during the training phase.
The notation "a ± b" represents the mean absolute error ± the standard deviation for each respective label and model.
}
\end{table*}

Table~\ref{tab:results_unsupervised} investigates the ability of our unsupervised models, trained with the unsupervised bottleneck, to predict labels for all models as a downstream task.
In this experiment, we used the models $\beta$-VAE and infoVAE, as well as real and mixed datasets.

As already demonstrated in Table~\ref{tab:results_main}, the unsupervised model $\beta$-VAE (U*) exhibits significantly higher errors than their semi-supervised or supervised counterparts (except the CNN encoder trained exclusively on ETC data).
Table~\ref{tab:results_unsupervised} shows consistent performance across all the unsupervised models and datasets we tested.
Additionally, adding ETC data to the training set (mixed data) seems to worsen the performance of the unsupervised models.
This is in contrast to the semi-supervised models, where the mixed data improves the performance.

\begin{table*}
    \caption{
    Utilizing outputs from unsupervised models for label prediction.
    }
    \label{tab:results_unsupervised}
    \centering
    \begin{subtable}{\textwidth}
        \centering
        \subcaption{Model Descriptions}
        \small
        \begin{tabular}{lllrl}
\toprule
 & Learning paradigm & Architecture & Bottleneck & Dataset \\
Legend descriptor &  &  &  \\
\midrule
betaVAE (U*) (1) & unsupervised & CNN-CNN & 128 & real \\
betaVAE (U) & unsupervised & CNN-ResNet & 128 & real \\
infoVAE (U) & unsupervised & CNN-ResNet & 128 & real \\
betaVAE (mix, U) & unsupervised & CNN-ResNet & 128 & mix \\
infoVAE (mix, U) & unsupervised & CNN-ResNet & 128 & mix \\
\bottomrule
\end{tabular}
    \end{subtable}
    \hfill
    \vspace{1mm}
    \begin{subtable}{\textwidth}
        \centering
        \subcaption{Intrinsic Labels}
        \small
        \begin{tabular}{lllll}
\toprule
 & \multirow{2}{*}{\shortstack[l]{$T_{\rm eff}$\\$(\mathrm{K})$}} & \multirow{2}{*}{\shortstack[l]{[M/H]\\($\mathrm{dex}$)}} & \multirow{2}{*}{\shortstack[l]{$\log(\mathrm{g})$\\($\mathrm{dex}$)}} & \multirow{2}{*}{\shortstack[l]{All\\(-)}} \\
 & & & & \\
\midrule
betaVAE (U*) (1) & 978.0 ± 13.0 & 0.1578 ± 0.0027 & 0.1798 ± 0.0025 & 1.548 ± 0.013 \\
betaVAE (U) & 986.0 ± 13.0 & 0.1579 ± 0.0027 & 0.1829 ± 0.0025 & 1.563 ± 0.013 \\
infoVAE (U) & 999.0 ± 13.0 & 0.1529 ± 0.0027 & 0.1817 ± 0.0025 & 1.562 ± 0.013 \\
betaVAE (mix, U) & 1017.0 ± 13.0 & 0.154 ± 0.0027 & 0.186 ± 0.0025 & 1.59 ± 0.013 \\
infoVAE (mix, U) & 990.0 ± 13.0 & 0.1648 ± 0.0026 & 0.1846 ± 0.0025 & 1.583 ± 0.013 \\
\bottomrule
\end{tabular}
    \end{subtable}
    \hfill
    \vspace{1mm}
    \begin{subtable}{\textwidth}
        \centering
        \subcaption{Extrinsic Labels}
        \small
        \begin{tabular}{lllll}
\toprule
 &  \multirow{2}{*}{\shortstack[l]{Radvel\\(\si{\kilo\meter\per\second})}}  & \multirow{2}{*}{\shortstack[l]{BERV\\(\si{\kilo\meter\per\second})}}  & \multirow{2}{*}{\shortstack[l]{Airmass\\(-)}} & \multirow{2}{*}{\shortstack[l]{All\\(-)}} \\
 & & & & \\
\midrule
betaVAE (U*) (1) & 31.07 ± 0.27 & 16.01 ± 0.096 & 0.181 ± 0.0019 & 1.5751 ± 0.0074 \\
betaVAE (U) & 32.15 ± 0.28 & 16.125 ± 0.097 & 0.1827 ± 0.0019 & 1.6056 ± 0.0075 \\
infoVAE (U) & 31.26 ± 0.28 & 16.077 ± 0.097 & 0.181 ± 0.0019 & 1.5811 ± 0.0075 \\
betaVAE (mix, U) & 31.72 ± 0.28 & 16.265 ± 0.1 & 0.1798 ± 0.0019 & 1.5935 ± 0.0076 \\
infoVAE (mix, U) & 31.81 ± 0.28 & 16.39 ± 0.1 & 0.1809 ± 0.0019 & 1.6017 ± 0.0076 \\
\bottomrule
\end{tabular}
\end{subtable}
\tablefoot{The column titled "All" uses NMAE from Eq.~\eqref{eq:NMAE} to summarize the mean absolute error across all normalized labels.
The notation "a ± b" represents the mean absolute error ± the standard deviation for each respective label and model.}
\tablebib{(1)~\citet{Nima_2021}.}
\end{table*}

\subsection{Machine learning simulation of spectra}
\label{sec:ML_simulation}
Our secondary goal is to create model spectra for given sets of physical parameters, as described in Sect.~\ref{sec:inference}.
This task can be approached using the isolated decoder or one of our semi-supervised models.
As the semi-supervised models can accommodate unsupervised information, the resulting simulations can reflect unknown statistical factors.
This difference is the main advantage over the isolated decoder or the traditional approach \citep[ETC;][]{ETC2}.
However, this advantage also complicates the evaluation, as it requires consideration of unknown statistical factors that by definition cannot be annotated.

We have already encountered these challenges, as shown in Fig.~\ref{fig:AE_bottleneck_RVIS}, where the number of unsupervised nodes is critical.
Too few unsupervised nodes $\mathbf{u}$ can compromise the precision of the label-informed nodes as they attempt to encode extra information unrelated to the actual labels.
In contrast, the excessively large size of $\mathbf{u}$ allows the model to reconstruct spectra from unsupervised nodes, without even utilizing label-informed nodes, thus breaking the connection between label-informed nodes and the output spectrum.
Thus, with real-world data, our ML models must balance high fidelity reconstruction (achieved with a large unsupervised bottleneck, as measured by the reconstruction error from Sect.\ref{sec:reconstruction}) against preserving cause-and-effect relationships in ML simulation (achieved with a small or constrained bottleneck, as measured by generative metrics from Sect.~\ref{sec:generative_metrics}). 
To navigate these inherent challenges of the ML simulation, we split the evaluation into three parts, each revealing a particular property.

First, we train stand-alone decoders (CNN and ResNet) for ML simulation using the loss function from Eq.~\eqref{eq:sim} and the ETC data. 
Then, we evaluate the reconstruction error for these decoders.
Because the ETC data are optimally represented by $\mathbf{l}$, we teach a decoder to map $\mathbf{l}$ to $\mathbf{s}$.
This allows us to analyze the behavior of the decoders without having to deal with unsupervised nodes or imprecise real data.
We determine the decoder with the smallest reconstruction error and use it in all our models, that is, we use real training data to train an AE model, and mixed training data to train AE and infoVAE model.

Second, we compute reconstruction errors for AEs, infoVAE, and reference model \citep{Nima_2021} on the real testing dataset.
This approach avoids the problem of unknown factors, as the encoders extract this information from the input spectrum.
By comparing the input and output spectrum, we can assess the implicit quality of the learned representation. 
This provides a global comparison between individual models.

Third, we examine the generative properties of AEs and infoVAE on the generative ETC dataset discussed in Sect.~\ref{sec:data}.
The generative properties measure how well the ML models emulate the cause-and-effect relationships and conclude our evaluation of ML simulation.

\subsubsection{Decoders and Exposure Time Calculator data}
\label{sec:simulations_results}
\begin{figure}[ht]
    \centering
    \includegraphics[width=0.5\textwidth]{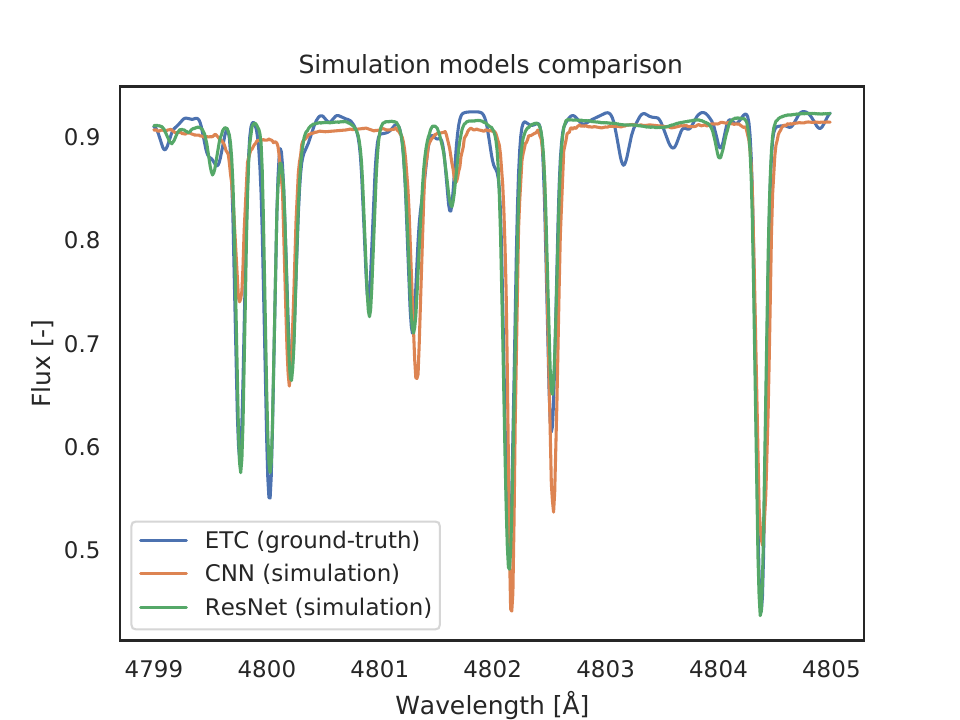}
    \caption{
        Comparison of reconstruction capabilities between CNN and ResNet models using the ETC dataset.
        The elevated error rate for the CNN model, as shown in Table~\ref{tab:simulations}, is attributed to missing absorption lines.
    }
    \label{fig:simulations_CNN_vs_ResNet}
\end{figure}

We start by evaluating the reconstruction error of individual decoders on the ETC dataset.
We utilize the ETC dataset because the unsupervised part of the bottleneck prevents us from evaluating this simplified ML simulation task on real data.
Table~\ref{tab:simulations} and Fig.~\ref{fig:simulations_CNN_vs_ResNet} present the results of our ML simulation experiments comparing CNN and ResNet decoders.
Both decoders were trained for 1,000 epochs using the Adam optimizer.
CNN (Table~\ref{tab:decoder_CNN}) used a learning rate of 0.0001, while ResNet (Table~\ref{tab:decoder_resnet}) used a learning rate of 0.001.

Table~\ref{tab:simulations} shows that the CNN decoder has a significantly higher reconstruction error $0.0139 \pm 0.0026$ compared to the ResNet decoder $0.0064 \pm 0.0011$.
We measure the error using MAE$'$, as defined in Eq.~\eqref{eq:MAE_rec}.
Although both decoders demonstrate proficiency in reconstructing continua, the CNN decoder frequently overestimates, underestimates, or misses the absorption lines.
This limitation is evident in Fig.~\ref{fig:simulations_CNN_vs_ResNet}, where, as a representative example, we show a narrow spectral region to highlight the individual characteristics of the spectra.
The figure illustrates a CNN decoder overestimating or underestimating almost all the lines and missing two lines around 4800 \text{\AA} and another around 4801 \text{\AA}.
In contrast, the ResNet decoder reconstructs the spectra with much greater accuracy.

This experiment has some limitations.
For example, the reconstruction loss does not take into account the causal impact of the labels.
This can lead to a situation where the model learns to ignore labels that have minimal impact on the output spectrum.
Nevertheless, this experiment is a prerequisite for a good simulation model and validates the decoder choice.

\begin{table*}[ht]
    \caption{
        Comparison of CNN and ResNet decoders for ETC data.
    }
    \centering
    \small
    \begin{tabular}{lllll}
\toprule
 & Model type & Learning paradigm & Architecture & Reconstruction error \\
Legend descriptor &  &  &  &  \\
\midrule
CNN (simulation) & decoder & supervised & CNN & 0.0139±0.0026 \\
ResNet (simulation) & decoder & supervised & ResNet & 0.0064±0.0011 \\
\bottomrule
\end{tabular}

    \label{tab:simulations}
\end{table*}

\subsubsection{Autoencoders and reconstruction results}
\begin{figure}[htbp]
    \centering
    \includegraphics[width=0.5\textwidth]{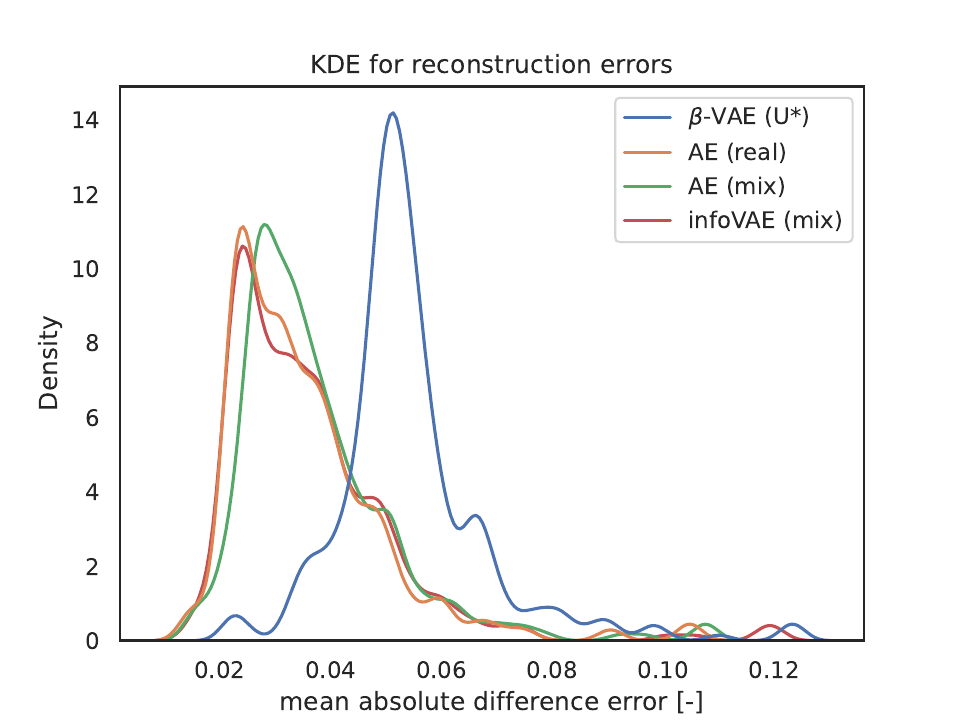}
    \caption{
        Kernel density plot for mean absolute error for reconstructions is shown. The KDE's bandwidth was set according to Scott's rule.
        The supervised infoVAE (CNN-ResNet) exhibits the lowest error, although its performance closely resembles that of AEs (CNN-ResNet).
        The unsupervised $\beta$-VAE (CNN-CNN), cited from \cite{Nima_2021}, shows the highest error, which can be attributed to its less complex decoder.
    }
    \label{fig:rec_error}
\end{figure}

\sloppy
The reconstruction error between the input and output spectra is a useful metric for both AEs and VAEs, as it can compensate for the lack of ground truth in unsupervised settings.
The results themselves demonstrate the efficiency of compression achieved by AEs or VAEs.
Figure~\ref{fig:rec_error} shows the distribution of reconstruction error (MAE$'$ from Eq.~\eqref{eq:MAE_rec}) for our two semi-supervised models (AE, infoVAE) and the $\beta$-VAE (U*) model from \cite{Nima_2021}.
The reconstruction error is evaluated using the real testing dataset.
The supervised models are excluded from this comparison as they do not produce reconstructions.

Overall, the error distributions are similar for AE (real), AE (mix), and infoVAE (mix).
The $\beta$-VAE with CNN decoder \citep{Nima_2021} has the poorest performance.
We hypothesize that AE (real) outperforms AE (mix) because the ResNet model is underfitting; thus, the added complexity from a mixed dataset results in poorer performance.
To validate this hypothesis, we would need to evaluate models more complex than ResNet, as listed in Table~\ref{tab:decoder_resnet}. We defer this evaluation to a future study.

The difference in spectra reconstruction between models is affected by the choice of the decoder, as already shown in Sect.~\ref{sec:simulations_results}.
While larger bottleneck size also helps with reconstruction \citep{Nima_2021}, it does so at the cost of generative properties (Sect.~\ref{sec:bottleneck_optimization}).

This experiment shares all the problems of the previous experiment.
To make matters worse, we cannot distinguish between the problematic encoder and decoder.
For example, the encoder may not learn meaningful representation, but the decoder can still reconstruct the spectra.
Conversely, even if the encoder learns meaningful representation, the decoder can still fail to reconstruct the spectra.
Similarly to the previous experiment, a good reconstruction error by itself does not translate to a good simulation model but is a necessary prerequisite.

\fussy

\subsubsection{Generative results}
Generative metrics from Sect.~\ref{sec:generative_metrics} are essential to evaluate the cause-and-effect connection between an individual spectral parameter and the output spectrum of our ML models.
These metrics directly target the shortcomings of the previous two experiments.
We examined the generative properties using both RVIS, as defined in Eq.\eqref{eq:rvis}, and GIS, as outlined in Eq.\eqref{eq:gis}.
We use boxplots to visualize the distribution of RVIS and GIS across various models.
A summary of our findings is presented in Table~\ref{tab:results_generative}.

Both generative metrics are based on reconstruction error, making it difficult to distinguish between continuum shift and incorrect line prediction.
Consequently, the metrics are mostly indicative.
The generative metrics can be interpreted as an additional reconstruction error, which is caused by constraining the ML models by the cause-and-effect relationship.

\begin{figure}
    \centering
    \includegraphics[width=0.5\textwidth]{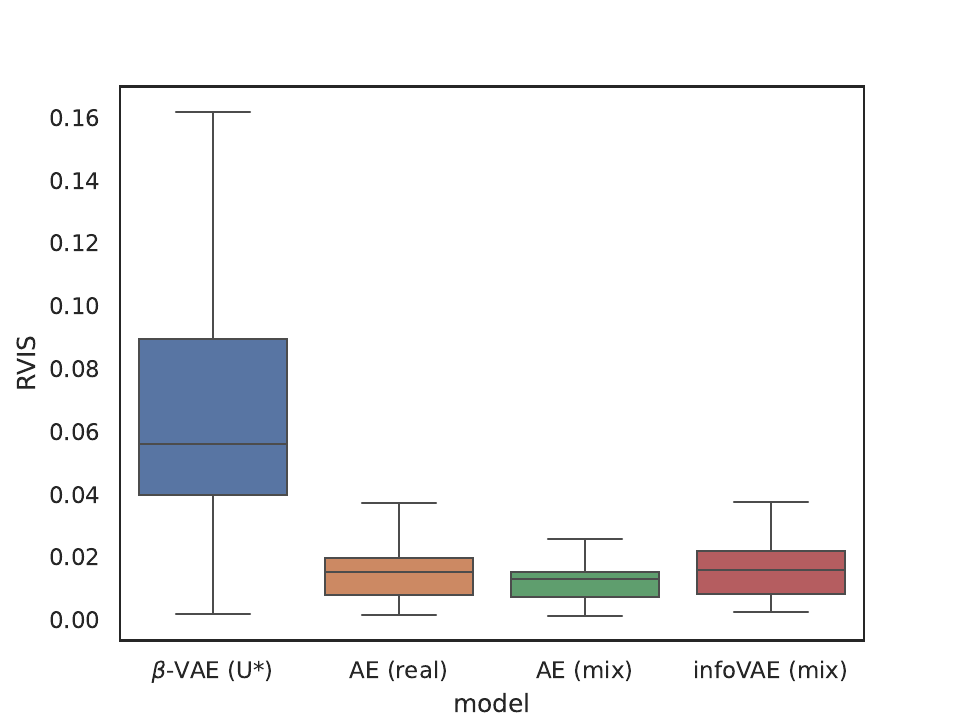}
    \caption{
    Radial velocity intervention score comparison between models boxplot.
    We do not include the supervised encoder because they do not provide reconstructions.
    The unsupervised model \citep{Nima_2021} is included since we can generate intervention over radial velocities for real data when concentrating on a wavelength range without tellurics.
    }
    \label{fig:RVIS_results}
\end{figure}

We investigated the RVIS for AE (real), AE (mix), and infoVAE (mix). Additionally, the unsupervised model from \cite{Nima_2021} is incorporated due to its capability, with proper calibration, to generate interventions over radial velocities for real data within telluric-free wavelength ranges.
The results in Fig.~\ref{fig:RVIS_results} and Table~\ref{tab:results_generative} show an average error close to 0.02 for semi-supervised models.
This is significantly better than the unsupervised model with a value of 0.063.

\sloppy
Even a minor discrepancy in radial velocity can lead to substantial reconstruction errors, as the entire spectrum is impacted.
Therefore, the relatively small RVIS observed for the semi-supervised model demonstrates it correctly captures radial velocity effects.

We investigated the GIS for AE (real), AE (mix), and infoVAE (mix). 
The unsupervised model from \cite{Nima_2021} was left out because it lacks the capability to generate new samples with predefined stellar parameters.
The focus of our investigation comprises labels such as temperature, metallicity, and gravity.

\begin{figure*}
    \centering
    \begin{subfigure}[b]{0.33\textwidth}
        \includegraphics[width=\textwidth]{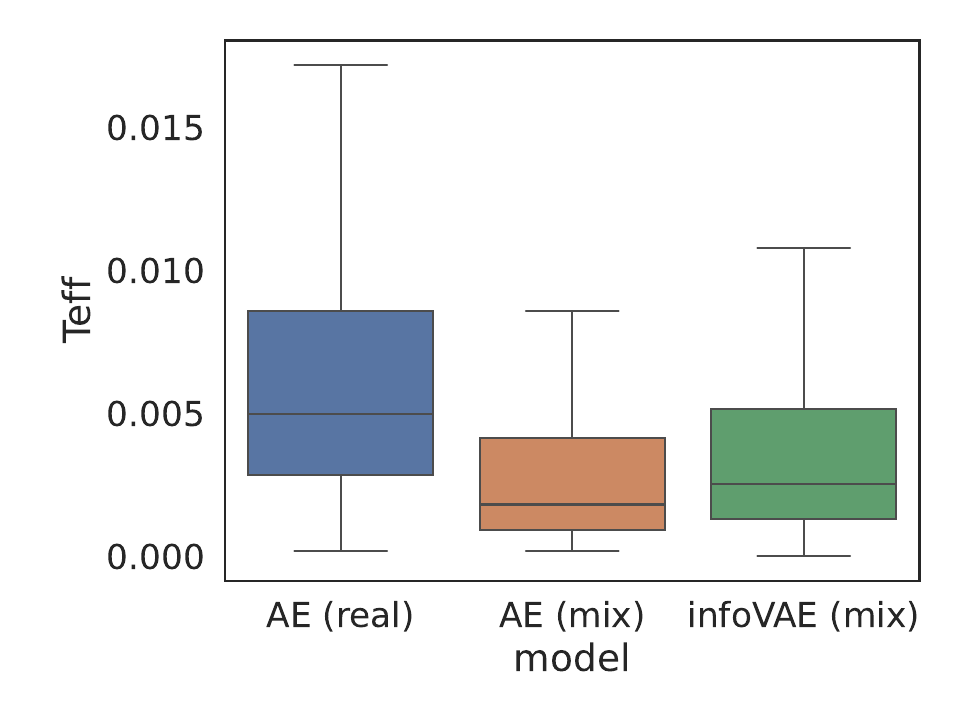}
        \caption{$T_\text{eff}$ [K]}
    \end{subfigure}
    \begin{subfigure}[b]{0.33\textwidth} \includegraphics[width=\textwidth]{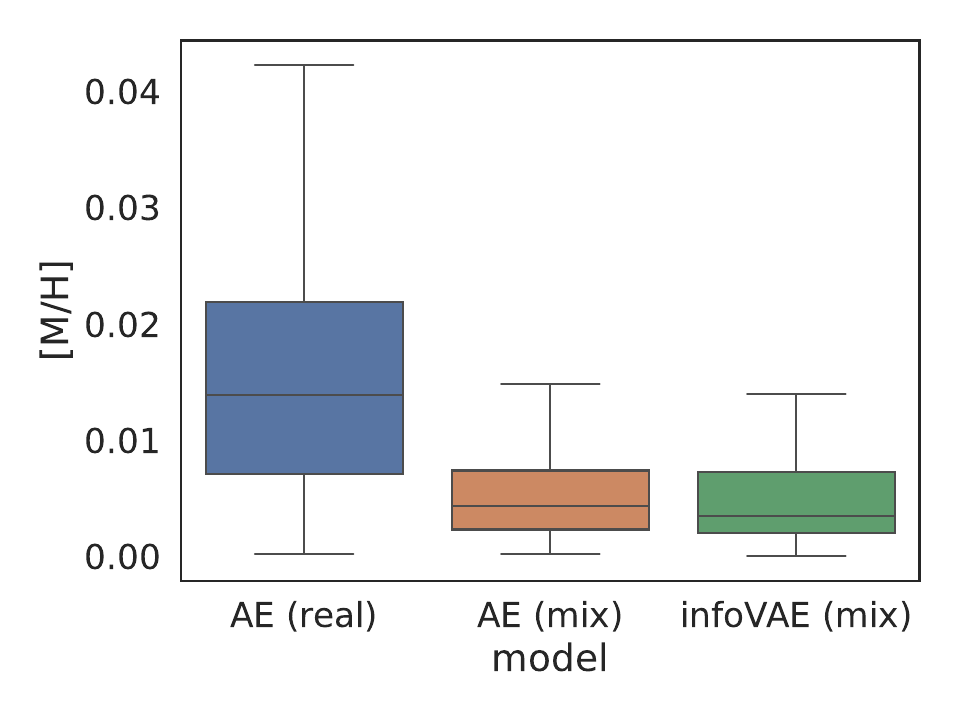}
        \caption{[M/H]}
    \end{subfigure}
    \begin{subfigure}[b]{0.33\textwidth}
        \includegraphics[width=\textwidth]{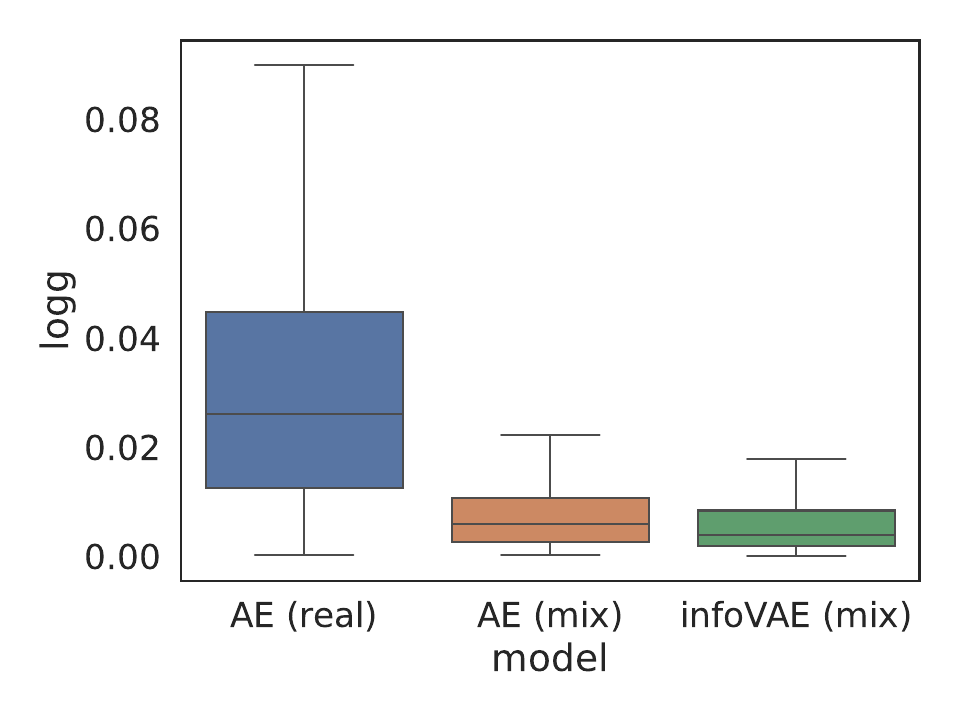}
        \caption{$\log g$}
    \end{subfigure}
    \caption{Boxplots showing GISs for our semi-supervised models.
    }
    \label{fig:GIS_results}
\end{figure*}

The GIS results in Fig.~\ref{fig:GIS_results} demonstrate excellent temperature modeling, with even the AE (real) model achieving an average error below 0.008.
The difference between the AE (real) and other models is even more pronounced for metallicity and surface gravity.
The results show that the reconstruction error due to physical parameters intervention is very small for all examined physical parameters.
This indicates that the investigated models can emulate the effects of the selected physical parameters.

\fussy
We train the AE (real) model exclusively on real data; therefore, its poorer performance could be attributed to an unfamiliar testing dataset. Alternatively, simulated data may have improved the generative outcomes of AE (mix) and InfoVAE (mix).
Given the similar performance for RVIS that is tested on the HARPS dataset, we suspect that the former scenario is likely.
More experimentation is required to determine the impact of simulated data on the generative properties.

The generative results suggest that once the ML hyperparameters are optimized, the distinction between AE and VAE becomes less significant.
Employing the label-aware paradigm remains crucial to enhance generative properties.
Moreover, these properties represent a valuable criterion for model selection, as described in  Sect.~\ref{sec:bottleneck_optimization}.

\sloppy
We presented three different experiments to evaluate a model's suitability for the ML simulation task.
First, we demonstrated a method for selecting the optimal decoder for reconstruction and identified the suitable ResNet decoder.
Second, we illustrated how to choose a model that can achieve high reconstruction fidelity for complete autoencoders, selecting the CNN-ResNet combination as a result.
Third, we introduced a novel approach to identify ML models that capture the cause-effect relationships between individual spectral parameters and the resulting spectrum, concluding that properly setting the ML hyperparameters causes the compared models to behave similarly.

\fussy

\begin{table}
    \caption{
    Generative results.
    }
    \label{tab:results_generative}
    \centering
    \small
    \begin{tabular}{lllll}
\toprule
& \multirow{2}{*}{\shortstack[l]{Radvel\\(-)}} & \multirow{2}{*}{\shortstack[l]{$T_{\rm eff}$\\(-)}} & \multirow{2}{*}{\shortstack[l]{[M/H]\\(-)}} & \multirow{2}{*}{\shortstack[l]{$\log(\mathrm{g})$\\(-)}} \\
& & & & \\
\midrule
$\beta$-VAE (U*) (1) & 0.063 & -- & -- & -- \\
AE (real) & 0.019 & 0.0078 & 0.017 & 0.036 \\
AE (mix) & 0.017 & 0.0041 & 0.0069 & 0.0077 \\
infoVAE (mix) & 0.02 & 0.0051 & 0.0065 & 0.0062 \\
\bottomrule
\end{tabular}
\tablefoot{Cells represent the generative mean absolute error for generative metrics.
Radvel is processed using RVIS from Eq.~\eqref{eq:rvis}.
The rest of the physical parameters are processed using GIS from Eq.~\eqref{eq:gis}.}
\tablebib{(1)~\citet{Nima_2021}.}
\end{table}

\subsection{Comparison of computational resources for different models}
\label{sec:discussion_time_and_memory}
Here, we describe how the time and memory requirements were evaluated.
We begin by discussing the setups of the time and memory experiment.
We are interested in the time and memory requirements of our ML models, additionally we analyzed the time requirements of the ETC simulator and the computation of intrinsic stellar SEDs with ATLAS9.
All experiments were restricted to use a single core of an AMD Ryzen™ 9 3900X, equipped with 64 GB of RAM.
For experiments that required GPU, we used a single NVIDIA GeForce RTX 3090, which features 24 GB of GDDR6X VRAM.

We evaluated the time requirements of our ML models by processing a single batch of 32 spectra and measuring the time per spectrum for CPU usage.
For GPU assessments, several such batches were processed.
The data used in this experiment were not loaded from disk, but were instead randomly generated on the RAM.

Memory requirements are measured by counting the number of ML parameters in each model.
This method provides a straightforward way to compare the memory demands of different models.
Such an assessment is valuable for evaluating the scalability of our models and determining their feasibility for deployment on various devices and datasets of different sizes.
The batch size, when processing large input data, significantly impacts practical memory requirements during inference.
Additionally, memory demands can vary substantially between the training and inference phases.
This variation is due to the need to store intermediate results during backpropagation.

Our time measurements showed that ATLAS9 takes 89 \si{\second} on average to generate a single spectrum using a single core.
Similarly, we measured ETC performance, and the processing time was 27 \si{\second}.
However, this time includes the time required to download the spectrum from the remote server.
As such, the ETC time measurements are not reliable.
Given that the ETC could be computed in principle extremely fast, we decided to exclude it from the timing comparison.
Instead, we consider the ATLAS9 time requirement as the baseline, as it is a prerequisite for ETC.

Table~\ref{tab:time_memory} presents the timing results and computational complexities for a range of machine learning models in inference mode, that is, processing new data using trained models.
Specifically, the table includes all autoencoders, encoders, and simulation models.

The time requirements for ML models, as shown in Table~\ref{tab:time_memory}, reveal that even the slowest model takes only 795 \si{\milli\second} on a single CPU core while generating the intrinsic stellar SED with ATLAS9 requires 89 \si{\second}.
Since this is a prerequisite for the ETC models, 89 \si{\second} represents the minimum time needed to generate a spectrum.
ML models can operate over a hundred times faster than ATLAS9 on identical CPUs.
Furthermore, ML models can be easily run on GPUs.
For instance, the CNN-CNN model from \cite{Nima_2021} takes just 0.386 \si{\milli\second} on a GPU, while the infoVAE CNN-ResNet model takes 3.970 \si{\milli\second}.

The ResNet model demonstrates a low reconstruction error in Table~\ref{tab:simulations} and efficient processing time in Table~\ref{tab:time_memory}, making it an appropriate option for significantly accelerating the computation of ETC (or SED).
On the other hand, the CNN model is even faster, with a reconstruction error that may be deemed acceptable for numerous applications.

In practice, the time requirements extend beyond just the runtime of the ML model due to I/O constraints.
Considering that each memory-optimized HARPS spectrum is approximately 1.2 MB, and we are loading batches of 32 spectra simultaneously, this translates to a data load of 38.4 MB per batch.
When SSDs offer read speeds ranging from 200 MB/s to 5500 MB/s, the loading time for one batch can vary between 0.007 and 0.192 seconds.
By contrast, conventional hard drives, which typically have read speeds ranging from 100 MB/s to 150 MB/s, would require approximately 0.256 to 0.384 seconds to complete the identical task.
These I/O times, especially when using slower hard drives, can become a dominating factor in the overall time efficiency of the system, often surpassing the runtime of the machine learning models themselves.
Additionally, loading large batches introduces overhead, potentially exacerbating the impact of I/O times on the total computation time.

The memory requirements for the ML models are presented in Table~\ref{tab:time_memory}.
In particular, the CNN-CNN model requires 7 million ML parameters, similar in scale to the CNN-ResNet, which has 14 million ML parameters.
An examination of CNN (simulation) and CNN (mix), the components of CNN-CNN, reveals that most ML parameter requirements stem from the larger bottleneck.

\sloppy
Our ML models are highly memory efficient.
For instance, 14 million ML parameters would roughly translate to 40 HARPS spectra.
Therefore, even hardware with modest memory and processing capabilities can efficiently infer labels or construct new spectra.
This presents a clear advantage over traditional approaches.

\fussy
\begin{table*}
\caption{
Comparative analysis of time and memory across models.
}
\label{tab:time_memory}
\centering
\small
\begin{tabular}{lllrlll}
\toprule
 & Model type & Architecture & Bottleneck & \multirow{2}{*}{\shortstack[l]{GPU\\(\si{\milli\second})}} & \multirow{2}{*}{\shortstack[l]{CPU\\(\si{\milli\second})}}  & \multirow{2}{*}{\shortstack[l]{Params\\(M)}}  \\
&  &  &  &  &  &  \\
Legend descriptor &  &  &  &  &  &  \\
\midrule
$\beta$-VAE (U*) (1) & $\beta$-VAE & CNN-CNN & 128 & 0.386 & 113.8 & 7.0 \\
infoVAE (mix) & infoVAE & CNN-ResNet & 32 & 3.970 & 779.6 & 14.0 \\
AE (mix) & AE & CNN-ResNet & 9 & 3.910 & 795.0 & 13.3 \\
CNN (mix) & encoder & CNN & 6 & 0.190 & 38.4 & 1.5 \\
CNN (simulation) & decoder & CNN & 6 & 0.217 & 76.3 & 1.7 \\
ResNet (simulation) & decoder & ResNet & 6 & 3.694 & 729.9 & 11.6 \\
\bottomrule
\end{tabular}
\tablefoot{The "GPU" column shows the time required to process a single spectrum using a GPU, measured as the time to process a batch divided by the batch size.
Similarly, the "CPU" column shows the results of the same experiment but using a CPU.
The "params" column represents the number of the ML parameters, expressed in millions.
References. }
\tablebib{(1)~\citet{Nima_2021}.}
 
\end{table*}

\section{Discussion}
\label{sec:discussion}
\sloppy
In this study, we systematically explored the influence of various techniques on label prediction and ML simulation for high-resolution spectra.
Our investigation spanned different learning paradigms, including supervised, unsupervised, and semi-supervised, as well as various architectures such as autoencoders and their variational counterparts.
Additionally, we delved into experimenting with decoder architectures and rigorously tuned ML hyperparameters to discover the most optimal variant while introducing novel metrics to evaluate ML simulations.

\fussy
The research examines HARPS spectra, real and simulated.
The proposed architecture can be easily adapted to process 1D spectra of any length, making it suitable for handling data of similar types.

The results presented in Table~\ref{tab:summary} demonstrate that label-aware learning is the most significant factor in determining the accuracy of the label prediction.
The differences within label-aware models appear to be more nuanced, and no single model emerged as the clear best.
Therefore, in applications where label prediction is the sole objective and a sufficient number of labels are available, it is sufficient to employ the supervised CNN model.
This approach minimizes computational and memory overhead while delivering excellent results.

Semi-supervised learning proved significant when we used as little as 1\% of the labels that we had available for real data, as shown in Table~\ref{tab:results_auxiliary}.
This can be useful when drawing conclusions from a limited amount of labeled data.
Stellar effective temperature, radial velocity, and BERV predictions benefit significantly from the inclusion of simulated data and semi-supervised techniques.
Most strikingly, temperature prediction improves twofold when comparing CNN 0.01 (real) with AE 0.01 (mix) in Table~\ref{tab:results_auxiliary}.
This suggests that labels with a strong impact on reconstruction benefit most from semi-supervision.
The strong connection between labels and spectra is further illustrated in Fig.~\ref{fig:lambda_labels}, where the label error demonstrates relative insensitivity to the label loss weight, suggesting that the network effectively recovers information from the spectra itself. 

Label prediction performance remained similar across various unsupervised models, as summarized in Table~\ref{tab:results_unsupervised}.
As anticipated, our supervised CNN model outperformed its unsupervised counterparts, reaffirming the value of supervision.
Our linear regression approach (downstream task from Sect.~\ref{sec:AE}) empirically highlights the inherent challenges of disentangled learning with unsupervised training.

Our findings show that ResNet decoders achieve lower reconstruction errors compared to the state-of-the-art CNNs, as reported by \cite{Nima_2021}.
This is consistent across both simulated (Table~\ref{tab:simulations}) and real datasets (Fig.~\ref{fig:rec_error}) when using bottlenecks of comparable sizes.
Qualitatively, ResNets significantly outperform CNNs in reconstructing absorption lines for both real and simulated data (Fig.~\ref{fig:simulations_CNN_vs_ResNet}).
This is a critical distinction, as absorption lines carry most of the relevant spectroscopic information.

\sloppy
As summarized in Table~\ref{tab:results_generative}, the GIS metric for temperature, metallicity, and surface gravity shows minimal reconstruction error due to intervention, with the highest error being at most 0.0078.
The RVIS is comparatively higher, approximately 0.02, for semi-supervised models.
However, the impact of radial velocity mismatch on the reconstruction error is significantly higher than any other label.

\fussy
The results for generative properties in Figs. \ref{fig:GIS_results} and \ref{fig:RVIS_results} showed a marginal difference between AEs and VAEs.
However, VAEs benefit from a probabilistic framework that allows for easier manipulation of the latent space.
Specifically, by assigning desired values to supervised nodes and sampling unsupervised nodes from the VAE's prior distribution, VAEs can generate spectra that not only mirror known labels $\mathbf{l}$ but also incorporate a degree of randomness to account for unknown factors $\mathbf{u}$, effectively sampling from  $f(\mathbf{u} | \mathbf{l}) = f(\mathbf{u}) = \mathcal{N}(\mathbf{0}, I)$, where $f$ denotes the probability distribution function.
The process critically depends on disentanglement, which separates unsupervised nodes both from each other and from supervised nodes, ensuring that the generative model can accurately reflect the underlying data structure.
Conversely, in AEs, the entanglement between nodes often becomes arbitrary, complicating the separation between supervised and unsupervised nodes.
This entanglement renders the conditional probability  $f(\mathbf{u} | \mathbf{l})$ effectively unknown, limiting the AE's capacity to generate novel observations by directly manipulating the latent space.

The RVIS metric emerged as an instrumental tool for tuning ML hyperparameters of all our semi-supervised models.
The RVIS experiment for semi-supervised AEs determined that the optimal size of the bottleneck is nine, as shown in Fig.~\ref{fig:AE_bottleneck_RVIS}.
Specifically, with seven supervised nodes, it suffices to add just two unsupervised nodes to optimize RVIS.
Expanding the bottleneck beyond nine elements to improve reconstruction comes at the cost of cause-and-effect relationship between supervised nodes and the output spectrum.
The ML model gradually loses the connection between the node representing the radial velocity and the true radial velocity.

We successfully tuned the VAE-based models using RVIS. 
We fixed the size of the bottleneck and focused on tuning ML hyperparameters $\lambda_{\mathbb{KL}}$ from Eq.~\eqref{eq:BVAE} and $\lambda_{\text{MI}}$ and $\lambda_\text{MMD}$ from Eq.~\eqref{eq:infoVAE}.
Typical tuning of these ML hyperparameters requires manual investigation of properties of each model, such as reconstruction error, mutual information, posterior collapse (unused nodes), and correlation between nodes \citep{Nima_2021,portillo}.
This manual investigation is necessary because we cannot a priori determine the optimal criteria.
The RVIS metric greatly simplified the process of tuning ML hyperparameters, as it provides a concrete and meaningful optimization objective.
We leave the analysis of the GIS metric as a model selection metric to future work, since it does not permit the use of real data for evaluation.

In this study, our models have demonstrated a significant advantage over standard approaches with respect to processing speed.
Our fast simulation models, in particular, have shown potential (Table~\ref{tab:simulations}) to accelerate the generation of ETC data, a benefit that is especially valuable for high-resolution spectra.
Although we observed a trade-off between reconstruction accuracy and processing time in machine learning models, our ResNet model, while slower, offers superior accuracy compared to the CNN model.
Given that our application does not require real-time solutions, the ResNet model is recommended where resources allow in order to leverage accuracy over speed.


\section{Summary and conclusions}
\label{sec:conclusions}
Our study successfully developed robust joint models that exhibit generative capabilities for both actual and simulated data.
These models are versatile tools well-suited for tasks including label prediction, real spectra encoding, anomaly detection, and the generation of highly realistic spectra that accurately reflect real-world scenarios.

In essence, our research underscores that while high-quality data and labels are essential for joint models with generative properties, the specific choice between AE and VAE remains noncritical.
Simulated data are a valuable resource, especially when real data are limited.
Such data ensure that the accuracy of label predictions is maintained thanks to their abundant availability.
Semi-supervised learning shows improvement in label prediction when access to labeled data is limited.

As illustrated in Tables~\ref{tab:results_main} and \ref{tab:results_auxiliary}, over a large fraction of the ML hyperparameter space the label-aware models provide an accuracy that is comparable to the most accurate ``traditional'' methods \citep{miller_2020}, making them competitive and attractive tools for quantitative spectroscopy.
In particular, this very good accuracy combined with the extremely fast execution times for inferring stellar parameters (see Table~\ref{tab:time_memory}; fractions of a second vs. minutes per spectra for most traditional methods) make the methods presented here an almost inescapable choice for high-throughout observations, such as the massive spectroscopic surveys that will be coming online in the near future (DESI-2, MSE, 4MOST, WEAVE, WSU), including if the science case requires near real-time follow up.

Our research has several limitations. First, we rely on the HARPS dataset as the primary evaluation tool, which limits the general applicability of our conclusions.
Second, we use normal distributions as priors in VAEs.
This introduces imprecision when handling known labels with different distributions and restricts the ability to discover labels within the unsupervised bottleneck that might also have non-normal distributions.
Furthermore, the inherent degeneracy of stellar spectra, where different combinations of temperature, metallicity, and surface gravity can produce similar-looking results, makes unimodal distributions ill suited for representing this complexity.

Additionally, while the VAE models for label prediction include both a mean vector and a standard deviation, we did not investigate the potential of the standard deviation as an uncertainty estimate, given its use for sampling and the KL penalty during training.
Instead, we opted to treat the output as point estimates derived from the mean vector, as this approach provided clearer interpretations of the predicted labels and avoided conflicts between the goals of uncertainty estimation, sampling, and regularization.

This research opens the door to several promising follow-up projects, including conducting a comprehensive analysis of the unsupervised elements $\mathbf{u}$ and extending our approach to diverse datasets, which can provide a broader understanding of our results.
Although we have explored the encoder architectures, we believe that there is more to uncover in this domain.
Future research can leverage our trained models to quickly simulate realistic samples, thereby enhancing the training of new models.
Moreover, we could focus on the probabilistic nature of VAEs, which by default provide likelihoods for their outputs and support non-Gaussian priors that could aid in discovering new labels.
We anticipate addressing some of these topics in a future paper.

\section*{Acknowledgment}
The first author (V. C.) extends thanks to the European Southern Observatory (ESO) for providing resources and support during his internship.
V. C. and R. S. acknowledge support by the OP VVV MEYS-funded project CZ.02.1.01/0.0/0.0/16\_019/0000765, "Research Center for Informatics" and the Czech Technical University internal grant SGS24/096/OHK3/2T/13.
We are grateful to the anonymous referee for carefully reading the manuscript and providing useful suggestions to improve it.
We are also grateful to the Ministry of Education, Youth, and Sports of the Czech Republic for supporting the first author's internship at ESO.

\bibliographystyle{aa}
\bibliography{my_bib} %

\begin{appendix}
\FloatBarrier
\section{Exposure Time Calculator dataset}
\label{sec:ETC}
Here, we describe how we used the ETC to simulate the HARPS instrument.
We began with with $N$ SED spectra from ATLAS9. For each spectrum, we sampled ETC settings (airmass, brightness, H20, \ldots) and fixed radial velocity to 0.
For each pair of SED and ETC settings, we performed the following:
\begin{enumerate}
    \item Process ETC settings and SED.
    Split the output into overlapping orders, where each order contains:
    \begin{itemize}
        \item target SED $\hat{S}$
        \item target signal $S$
        \item sky signal $B$
        \item atmospheric transmission $a$
        \item blazing $b$
        \item dispersion $d$ [nm/px] (JSON output is actually [m/px])
        \item readout noise RON $e^-$/bin
        \item number of detector integrations NDIT
        \item dark current DARK
        \item detector integration time DIT
        \item number of spatial pixels $N_\text{spat}$
        \item number of spectral pixels $N_\text{spec}$        
    \end{itemize}
    \item Define $N_\text{exp} \equiv \text{NDIT}$ and $T_\text{exp} \equiv \text{DIT} \cdot \text{NDIT}$. 
    \item Denote $h = bd$. (We removed $h$
    to gain overlapping orders.)
    \item Denote $n = N_\text{spat} N_\text{spec} N_\text{exp} (T_\text{exp} \cdot \text{dark} + \text{RON}^2)$.
    \item Denote the concatenation of even orders with subindex $A$.
    \item Denote the concatenation of odd orders with subindex $B$.    
    \item Compute the total throughput as $T_A = S_A/\hat{S}_A$ and $T_B = S_B/\hat{S}_B$.
    \item Save $\hat{S}$, $B_A$, $B_B$, $T_A$, $T_B$, $h_A$, $h_B$, $n_A$, $n_B$. 
    For a single instrument, $n_{A, B}$ is fixed, and $\hat{S}$ is considered a continuous signal without the even or odd splitting.
    Therefore, we have a sevenfold increase in memory.
    \label{item:precompute} 
\end{enumerate}

Then, at runtime (given the precomputed variables from item~\ref{item:precompute} and an arbitrary radial velocity $v$) we perform the following steps to get the simulated flux.
\begin{enumerate}
    \item Load $\hat{S}$, $B_A$, $B_B$, $T_A$, $T_B$, $h_A$, $h_B$, $n_A$, $n_B$.
    \item Compute shifted SED as $\hat{S}_v$.
    \item Compute $S_A = T_A \hat{S}_v$ and $S_B = T_B \hat{S}_v$.
    \item Define the output signal $S$ as
    \begin{equation}
        S = 
        \begin{cases}
            \left(\frac{S_A}{h_A} + \frac{S_B}{h_B}\right)\frac{1}{2}, & \text{iff A and B overlaps} \\
            \frac{S_*}{h_*}, & \text{otherwise}
        \end{cases}
        \label{eq:signal}
    \end{equation}
    \item Define the variance $\sigma^2$ as
    \begin{equation}
        \sigma^2 = 
        \begin{cases}
            \left(\frac{S_A + B_A + n_A}{h_A^2} + \frac{S_B + B_B + n_B}{h_B^2} \right)\frac{1}{2^2}, & \text{iff A and B overlaps} \\
            \frac{S_* + B_* + n_*}{h_*^2}, & \text{otherwise}
        \end{cases}
        \label{eq:sigma}
    \end{equation}
    \item Sample the final output (per pixel) from $\mathcal{N}(S, \sigma)$.
\end{enumerate}

\FloatBarrier
\section{Architectures}
\label{sec:archs}
The architectures of the CNNs and ResNets used in this study are detailed below.
This section is intended to aid a reimplementation of our work.

Convolutional neural networks encoder architecture: As outlined in \cite{Nima_2021} and depicted in Tab.~\ref{tab:encoder_CNN}, each encoder layer consists of a 1D convolution with a dilation of 1 and a Leaky Rectified Linear Unit (Leaky ReLU) activation function with a negative slope of $0.1$.
The hidden units for each layer are defined by its ``chanels\_in'' and ``chanels\_out'' parameters, with the convolutional kernel's size and stride varying per layer. For instance, layer 4 has chanels\_in = 16 and chanels\_out = 32. 

Decoder architecture: The decoder in Tab.~\ref{tab:decoder_CNN} employs ConvTranspose1D layers combined with LeakyReLU activation functions (0.1 slope). Similar to the encoder, each layer's configuration is defined by its chanels\_in and chanels\_out, along with the convolutional kernel's size and stride.

\begin{figure}[ht]
     \centering
\begin{tikzpicture}[
    block/.style={rectangle, draw, fill=blue!20, text width=2.5em, text centered, rounded corners, minimum height=2em},
    line/.style={draw, -{Latex}},
    node distance=0.3cm
]
    \node[block] (input) {In};
    \node[block, below=of input] (inorm1) {IN1};
    \node[block, below=of inorm1] (act1) {A1};
    \node[block, below=of act1] (conv1) {C1};
    \node[block, right=of conv1] (upsample) {Ups};  
    \node[block, below=of conv1] (inorm2) {IN2};
    \node[block, below=of inorm2] (act2) {A2};
    \node[block, below=of act2] (conv2) {C2};
    \node[block, below=of conv2] (add) {Add};
    \node[block, below=of add] (output) {Out};

    \path[line] (input) -- (inorm1);
    \path[line] (inorm1) -- (act1);
    \path[line] (act1) -- (conv1);
    \path[line] (conv1) -- (inorm2);
    \path[line] (inorm2) -- (act2);
    \path[line] (act2) -- (conv2);
    \path[line] (conv2) -- (add);
    \path[line] (add) -- (output);
    \path[line] (input) -| (upsample);
    \path[line] (upsample) |- (add);
\end{tikzpicture}
\caption{Diagram of the PreActTransResNetBlock, illustrating the flow from ``In'' (input) through various components — ``IN'' (instance normalization), ``A'' (activation), ``C'' (transpose convolution), ``Ups'' (upsampling), ``Add'' (addition) — to ``Out'' (output).}
\label{fig:preacttransresnetblock}

\end{figure}
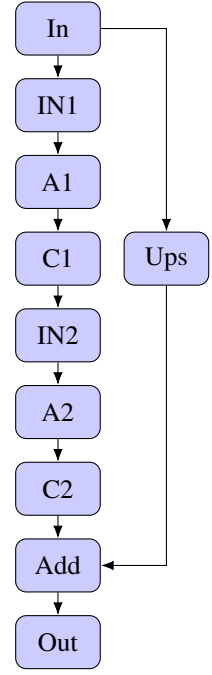

\sloppy
PreActTransResNetBlock: This block in Fig.~\ref{fig:preacttransresnetblock} accepts input "In," which is the output of the preceding block, and outputs "Out," which feeds to the next block or is the final output. We use the following components: instance normalization (IN), activation (A), ConvTranspose1D (C), and upsampling (Ups).
The block is defined by the ``C'' configuration, which is the basic building block of the CNN encoder.
``C1'' has parameters ``in\_chanels'' = chanels\_in, ``out\_chanels'' = chanels\_in, ``kernel\_size'' = 3, ``stride'' = 2, ``dilation'' = 1, while ``C2'' has in\_chanels = chanels\_in, out\_chanels = chanels\_out, kernel\_size = 3, stride = 1, dilation = 1.
These blocks allow for a modular building of the entire ResNet decoder.

\fussy
The residual network decoder architecture: Each block of the ResNet decoder in Tab.~\ref{tab:decoder_resnet} is a PreActTransResNetBlock from Fig.~\ref{fig:preacttransresnetblock}.
For instance, layer 4 has chanels\_out = 256 and chanels\_in = 512. This architecture ensures efficient and effective feature translation from encoded to decoded representations.

These architectures are pivotal in our model's ability to accurately process and reconstruct complex data structures.

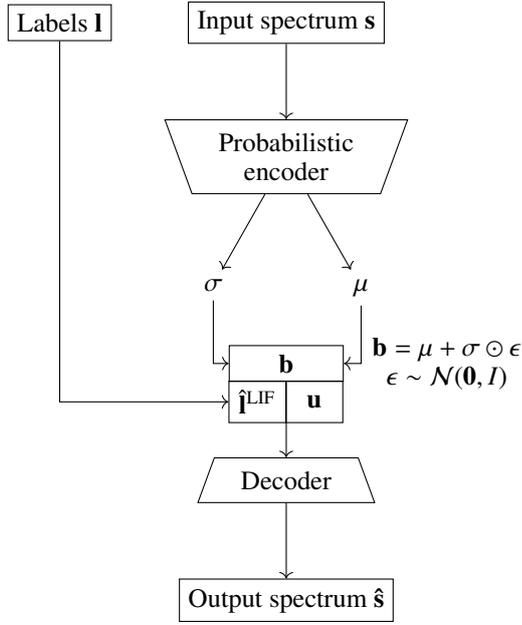
\begin{figure}
    \centering
    \setkeys{Gin}{width=1.0\linewidth}
            
    \begin{tikzpicture}[node distance=1cm, auto]
        \node[draw] (input_spectrum) {Input spectrum $\mathbf{s}$};
        \node[trapezium, 
        trapezium left angle=110, 
        trapezium right angle=110, 
        draw, 
        below=of input_spectrum,
        inner xsep=10pt, 
        inner ysep=5pt,  
        align=center] (encoder) {Probabilistic\\encoder};
        \node[below right=1cm and -1.7cm of encoder] (sigma) {$\mathbf{\sigma}$};
        \node[below left=1cm and -1.7cm of encoder] (mu) {$\mathbf{\mu}$};        
        \node[below=2cm of encoder, draw, minimum width=1.5cm] (sample) {$\mathbf{b}$};
        \node[right=0.25cm of sample, align=center] (reparametrization) {$\mathbf{b} = \mathbf{\mu} + \mathbf{\sigma} \odot \mathbf{\epsilon}$ \\ $\mathbf{\epsilon} \sim \mathcal{N}(\mathbf{0}, I)$};
        \node[below left=-0.05em and -0.76cm of sample, draw, minimum width=0.75cm, minimum height=0.55cm] (l) {$\hat{\mathbf{l}}^{\text{LIF}}$};
        \node[below right=-0.05em and -0.77cm of sample, draw, minimum width=0.75cm,minimum height=0.55cm] (u) {$\mathbf{u}$};
        \node[trapezium, 
        trapezium left angle=70, 
        trapezium right angle=70, 
        draw, 
        below=of sample,
        inner xsep=10pt, 
        inner ysep=5pt,  
        align=center] (decoder) {Decoder};   
        \node[below=of decoder,draw] (output_spectrum) {Output spectrum $\mathbf{\hat{s}}$};


        \draw[->] (input_spectrum) -- (encoder);
        \draw[->] (encoder) -- (sigma);
        \draw[->] (encoder) -- (mu);
        \draw[->] (mu) |- (sample);
        \draw[->] (sigma) |- (sample);
        \draw[->] (sample) -- (decoder);
        \draw[->] (decoder) -- (output_spectrum);

        \node[draw, left=of input_spectrum, align=center] (label_block) {Labels $\mathbf{l}$};
        \draw[->] (label_block) |- (l);

    \end{tikzpicture}
    \caption{Illustration of a semi-supervised VAE architecture. Labels $\mathbf{l}$ provide bottleneck supervision (injection).
    Input spectrum $s$ is encoded by probabilistic $q$ into an isotropic Gaussian $\mathcal{N}(\boldsymbol{\mu}, \boldsymbol \sigma)$.
    Latent representation is obtained from this distribution using the reparameterization trick from \cite{VAE}, where $\odot$ denotes the element-wise product.
    We apply the probabilistic decoder $p$ to the latent representation to obtain output spectrum $\mathbf{\hat{s}}$.    
    }
    \label{fig:illustration}
\end{figure}

\begin{table}
\caption{
    Encoder layers based on 1D convolution from \cite{Nima_2021}.
}
\centering
\small
\begin{tabular}{cccc}
        \toprule
        Layer & Hidden Units & Kernel Size & Stride \\
        \midrule
        1 & 16 & 7 & 2 \\
        2 & 16 & 5 & 2 \\
        3 & 16 & 5 & 2 \\
        4 & 32 & 3 & 2 \\
        5 & 32 & 3 & 2 \\
        6 & 32 & 3 & 2 \\
        7 & 64 & 3 & 2 \\
        8 & 64 & 3 & 2 \\
        9 & 64 & 3 & 2 \\
        10 & 128 & 3 & 2 \\
        11 & 128 & 3 & 2 \\
        12 & 128 & 3 & 2 \\
        13 & 256 & 3 & 2 \\
        14 & 512 & 3 & 2 \\
        15 & 512 & 3 & 1 \\
        \bottomrule
\end{tabular}
\tablefoot{All layers use Leaky ReLU (slope $0.1$) and dilation of 1.
chanels\_in and chanels\_out specify hidden units for each layer, for example, layer 4: chanels\_in = 16, chanels\_out = 32.}
\label{tab:encoder_CNN}
\end{table}

\begin{table}[ht]
    \centering
    \caption{
        Decoder architecture from \cite{Nima_2021}. 
    }
    \label{tab:decoder}
    \small
    \begin{tabular}{cccc}
        \toprule
        Layer & Hidden Units & Kernel Size & Stride \\
        \midrule
        1 & 512 & 3 & 1 \\
        2 & 256 & 4 & 2 \\
        3 & 128 & 4 & 2 \\
        4 & 128 & 4 & 2 \\
        5 & 128 & 4 & 2 \\
        6 & 64 & 4 & 2 \\
        7 & 64 & 4 & 2 \\
        8 & 64 & 4 & 2 \\
        9 & 32 & 4 & 2 \\
        10 & 32 & 4 & 2 \\
        11 & 32 & 4 & 2 \\
        12 & 16 & 4 & 2 \\
        13 & 16 & 4 & 2 \\
        14 & 16 & 4 & 2 \\
        \bottomrule
    \end{tabular}
    \label{tab:decoder_CNN}
    \tablefoot{
    All layers use ConvTranspose1D with LeakyReLU(0.1) as the activation function.
        Hidden units for layer $i$ are its chanels\_in and chanels\_out for layer $i + 1$.
        For example, layer 6 has chanels\_in = 128 and chanels\_out = 64.
        }
\end{table}

\begin{table}[ht]
\caption{Architecture of the decoder ResNet. }
\centering
\small
\begin{tabular}{cc}
\hline
Block Number & Hidden Units \\
\hline
1 & 512 \\
2 & 512 \\
3 & 512 \\
4 & 256 \\
5 & 256 \\
6 & 256 \\
7 & 128 \\
8 & 128 \\
9 & 64 \\
10 & 64 \\
11 & 32 \\
12 & 32 \\
13 & 16 \\
14 & 16 \\
15 & 16 \\
\hline
\end{tabular}
\tablefoot{
Each block is a PreActTransResNetBlock from Fig.~\ref{fig:preacttransresnetblock} with a ReLU activation function.
Hidden units for layer $i$ are its chanels\_out and chanels\_in for layer $i + 1$. That is, layer 4 has chanels\_out = 256 and chanels\_in = 512.
}
\label{tab:decoder_resnet}
\end{table}

\FloatBarrier
\section{Variational autoencoder details}
\label{sec:vaes}
Here, we are going to describe essential concepts of VAEs, which are important to fully understand our text.
We start with introducing the original VAEs \citep{VAE} and connecting it to our approach.
Then, we describe the InfoVAE \citep{infoVAE} and how we use it in our work.

\subsection{Variational autoencoders}
\label{sec:VAE_basics}
Variational autoencoders are a powerful class of generative models that facilitate the learning of latent representations of input data. Their significance lies in their ability to handle complex data distributions by combining neural networks and probabilistic graphical models.

The core challenge that VAEs address is the intractability of direct likelihood maximization in complex data distributions. Traditional methods of computing likelihoods are often computationally expensive and impractical for high-dimensional data. VAEs circumvent this problem by introducing the concept of the Evidence Lower Bound (ELBO), as proposed in the seminal work on VAEs \citep{VAE}.

The Evidence Lower Bound is the objective for VAEs, serving as an attainable lower bound to the intractable log-likelihood of the observed data
\begin{equation}
\mathcal{L}(\theta, \phi, \mathbf{s}) = -\mathbb{E}_{\mathbf{b} \sim q_\phi(\mathbf{b} \mid \mathbf{s})} \left[ \log p_\theta(\mathbf{s} \mid \mathbf{b}) \right] + D_{\mathbb{KL}}(q_\phi(\mathbf{b} \mid \mathbf{s}) || p(\mathbf{b}))\,,
\label{eq:ELBO}
\end{equation}
where $\theta$ and $\phi$ represent the parameters of the decoder and encoder networks, respectively.
The term $\mathbf{s}$ denotes the input data, and $\mathbf{b}$ is the latent variable.
The ELBO decomposes into two components.

The first term in Eq.~\eqref{eq:ELBO}, the reconstruction loss, is the expected negative log-likelihood of the observed data, given the latent representation.
This term encourages the model to accurately reconstruct the input data from its latent representation.

The second term in ELBO is the KL divergence between the approximate posterior $q_\phi(\mathbf{b} \mid \mathbf{s})$ and the prior distribution $p(\mathbf{b})$.
This divergence measures the difference between the two probability distributions, quantifying the amount of information lost when using $q_\phi$ to approximate $p(\mathbf{b})$.
It is defined as
\begin{equation}
D_{\mathbb{KL}}(q(\mathbf{b}) || p(\mathbf{b})) = \mathbb{E}_{\mathbf{b} \sim q(\mathbf{b})} \left[ \log \frac{q(\mathbf{b})}{p(\mathbf{b})} \right].
\label{eq:KL}
\end{equation}

While the standard VAE, as described earlier, focuses on a balanced reconstruction of input data and latent space regularization, the $\beta$-VAE \citep{Beta_VAE} introduces a more nuanced control over this balance.
By incorporating the hyperparameter $\beta$ that scales KL divergence, the $\beta$-VAE allows for a deliberate emphasis on either aspect, depending on the specific requirements of the task at hand.
This flexibility is particularly useful in scenarios where disentangling the latent features is more critical than achieving high-fidelity reconstruction, offering a tailored approach to complex data representation challenges.

\sloppy
It can be shown that the reconstruction loss from Eq.~\eqref{eq:rec} is equivalent to the first term $\mathbb{E}_{\mathbf{b} \sim q(\mathbf{b})} \left[ \log p_{\theta}(\mathbf{s} \mid \mathbf{b}) \right]$ in Eq.~\eqref{eq:ELBO}, provided certain assumptions are met.
The first term is estimated using the Monte Carlo approach, often with the sample size one \citep{VAE}.
Hence $\mathbb{E}_{\mathbf{b} \sim q(\mathbf{b})} \left[ \log p_{\theta}(\mathbf{s} \mid \mathbf{b}) \right]$ is simplified to $ \log p_{\theta}(\mathbf{s} \mid \mathbf{\hat{b}})$, where $\mathbf{\hat{b}} \sim q(\mathbf{b})$.
Assuming $p_{\theta}$ is a Laplacian distribution, the log-likelihood term $\log p_{\theta}(\mathbf{s} \mid \mathbf{\hat{b}})$ simplifies to $-\frac{|\mathbf{s} - \mathbf{\mu}(\mathbf{\hat{b}})|}{\mathbf{s}'} - \log(2\mathbf{s}')$, where $\mathbf{\mu}(\mathbf{\hat{b}})$ represents the mean vector and $\mathbf{s}'$ the scale vector (we note that this $\mathbf{s}'$ is distinct from the spectral data $\mathbf{s}$).
If we fix $\mathbf{s}'$ to a vector of ones, this expression reduces to a standard L1 norm, since the division by $\mathbf{s}'$ becomes trivial and the $-\log(2\mathbf{s}')$ term becomes a constant. Alternatively, setting $\mathbf{s}'$ to any constant value leads to a scaled L1 norm. 

\fussy
We note that if we treat the Laplace scale $\mathbf{s}'$ as a hyperparameter, the original VAE is equivalent to $\beta$-VAE, where $\lambda_\beta = \mathbf{s}'$.
This can be valuable as modifications to the ELBO can cause it to lose its probabilistic interpretation as a strict evidence lower bound.
This point of view opens up probabilistic interpretation even for the $\beta$-VAE objective.

\subsection{InfoVAE}
\label{sec:infoVAE_basics}
InfoVAE is one of many modifications of VAEs.
Similar to $\beta$-VAE, it allows us to control the balance between reconstruction and regularization.
Unlike $\beta$-VAE, it can achieve greater regularization without risking underutilization of the bottleneck.

This phenomenon, typically known as posterior collapse, is a common problem in VAEs and its derivatives.
It occurs when the KL term dominates ELBO, causing the posterior distribution $q^p_\phi(\mathbf{b} \mid \mathbf{s})$ to collapse to the prior distribution $p_b(\mathbf{b})$  \citep[pp.~796--797]{Murphy_23}.

InfoVAE solves this problem by introducing an additional term that encourages the latent space to be informative with respect to the input spectrum.
The semantic meaning of this VAE modification is shown through the following equation:
\begin{align}
L_{\text{infoVAE}}(\theta, \phi) &=
-\lambda_\mathbb{KL} D_{\mathbb{KL}} (q^p_\phi(\mathbf{b}) || p_b(\mathbf{b})) \nonumber
\\&\mkern-80mu
-\mathbb{E}_{\mathbf{b} \sim q^p_\phi(\mathbf{b})} \Bigl[
D_{\mathbb{KL}}(q^p_\phi(\mathbf{s} \mid \mathbf{b}) || p_\theta(\mathbf{s} \mid \mathbf{b}))
\Bigr]
+ \lambda_\text{MI} I_{q^p_\phi(\mathbf{s}, \mathbf{b})} (\mathbf{s}, \mathbf{b}),
\label{eq:InfoVAE_semantic}
\end{align}
where $I_{q^p_\phi(\mathbf{s}, \mathbf{b})} (\mathbf{s}, \mathbf{b})$ is the mutual information between $\mathbf{s}$ and $\mathbf{b}$, $\lambda_\mathbb{KL}$ represents the weight of the KL term (in a form that prevents posterior collapse), and $\lambda_\text{MI}$ represents the weight assigned to the mutual information term.
Notice that $q^p_\phi(\mathbf{b})$ in the first term ${\mathbb{KL}} (q^p_\phi(\mathbf{b}) || p_b(\mathbf{b}))$ is no longer conditioned on the input spectrum $\mathbf{s}$.
The first term brings us closer to the goal of having $q^p_\phi$ match the priors $p_b$, but requires a difficult computation of the marginal distribution $q^p_\phi(\mathbf{b}) = \int q^p_\phi(\mathbf{b}, \mathbf{s}) \, \mathrm{d} \mathbf{s}$.

We have two degrees of freedom in the objective Eq.~\eqref{eq:InfoVAE_semantic}.
We can adjust $\lambda_\mathbb{KL}$ to balance the disentanglement requirement directly, while $\lambda_\text{MI}$ determines to what degree the latent space is informative with respect to the input spectra, thus preventing posterior collapse.

The main text contains a more computationally friendly loss function in Eq.~\eqref{eq:infoVAE_equivalent}.
This loss function is equivalent to that in Eq.~\eqref{eq:InfoVAE_semantic} up to a constant (see the appendix in \citealt{infoVAE} for a full derivation).
This form allows us to use the KL term from Eq.~\eqref{eq:BVAE}, and we do not have to compute the mutual information term explicitly.

The loss function Eq.~\eqref{eq:infoVAE_equivalent} generalizes the loss function of both VAE and $\beta$-VAE.
Notice that we can achieve VAE by setting $\lambda_\text{MI} = 0$ and $\lambda_\text{MMD} = 1$.
This implies that VAE can be considered a specific instance of InfoVAE where the mutual information between the latent space and input spectra is not explicitly optimized.
Further notice that we can achieve equivalency with $\beta$-VAE by setting $\lambda_\text{MI} = 1 - \lambda_{\mathbb{KL}}$ and $\lambda_\text{MMD} = 1 - \lambda_\text{MI}$.

The computation of $D_{\mathbb{KL}}\left(q^p_\phi(\mathbf{b}) \| p(\mathbf{b})\right)$ is the only remaining problem in Eq.~\eqref{eq:infoVAE_equivalent}. The work \cite{infoVAE} proposes using Maximum-Mean Discrepancy (MMD) to approximate the KL divergence.

The Maximum-Mean Discrepancy is a statistical technique in machine learning and statistics that measures the dissimilarity between two probability distributions \citep{MMD}.
It is particularly useful when we cannot compare the distributions directly or when they are not known explicitly.
MMD compares the means of samples drawn from the two distributions when mapped to a higher-dimensional feature space, typically using a kernel function (kernel trick).
The basic idea is that if the means are close in the feature space, then the distributions are likely to be similar.
A full theoretical explanation is provided in \cite{MMD}.
We can compute the MMD that approximates the term $D_{\mathbb{KL}}\left(q^p_\phi(\mathbf{b}) \| p(\mathbf{b})\right)$ in Eq.~\eqref{eq:infoVAE_equivalent} as follows
\begin{align}
D_{\text{MMD}}(p_{b}, q^p_{\phi}) &= \mathbb{E}_{\mathbf{b} \sim p_b(\mathbf{b})} \mathbb{E}_{\mathbf{b}' \sim p_b(\mathbf{b}')} \left[ k(\mathbf{b}, \mathbf{b}') \right] \nonumber \\ &- 2 \mathbb{E}_{\mathbf{b} \sim p_b(\mathbf{b})} \mathbb{E}_{\mathbf{b}' \sim q^p_\phi(\mathbf{b}' \mid \mathbf{s})} \left[ k(\mathbf{b}, \mathbf{b}') \right] \nonumber \\ &+ \mathbb{E}_{\mathbf{b} \sim q^p_\phi(\mathbf{b} \mid \mathbf{s})} \mathbb{E}_{\mathbf{b}' \sim q^p_\phi(\mathbf{b}' \mid \mathbf{s})} \left[ k(\mathbf{b}, \mathbf{b}') \right],
\label{eq:MMD}
\end{align}
where $k(\mathbf{b}, \mathbf{b}')$ is an arbitrary positive definite kernel function, such as a Gaussian kernel.

\FloatBarrier
\section{Analysis of $\lambda_\text{lab}$ hyperparameter}
\label{sec:lambda_lab}

\begin{figure}[htbp]
    \centering
    \includegraphics[width=0.5\textwidth]{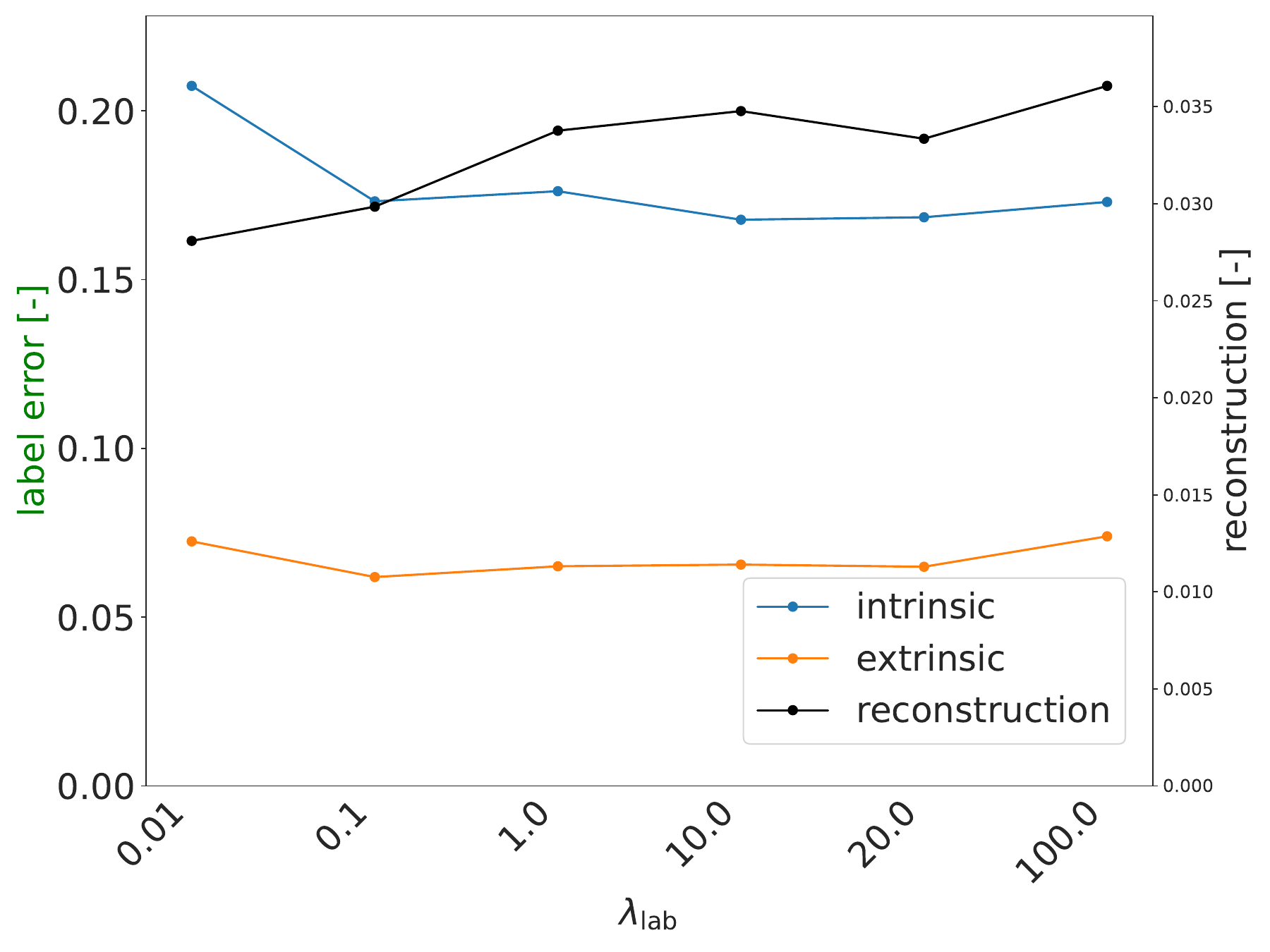}
    \caption{
       Impact of hyperparameter $\lambda_\text{lab}$ from Eq.~\eqref{eq:AE} on label error and reconstruction error.
       The left x-axis corresponds to the intrinsic and extrinsic curves, while the right x-axis corresponds to the reconstruction curve.
    }
    \label{fig:lambda_labels}
\end{figure}

We investigated the impact of the hyperparameter $\lambda_\text{lab}$ from Eq.~\eqref{eq:AE}, which balances the reconstruction loss and label loss for AEs.
We experimented with values ranging from $0.01$ to $100$ and observed the effects on both reconstruction loss and label loss, as shown in Fig.~\ref{fig:lambda_labels}.
All models were trained on a mixture of real and ETC data.
In the graph, the x-axis represents the values of $\lambda_\text{lab}$, while the y-axis on the left shows the normalized label prediction error, and the y-axis on the right shows the reconstruction error.
The blue curve in Fig.~\ref{fig:lambda_labels} displays errors for intrinsic labels (temperature, gravity, and metallicity), the red curve shows errors for extrinsic labels (radial velocity, barycentric Earth radial velocity, and airmass), and the black curve represents the reconstruction error.

The range of errors for both reconstruction and label loss is relatively small, indicating insensitivity to this parameter.
Label loss is especially insensitive to changes within the interval $[0.1, 20]$. Therefore, for simplicity, we set this hyperparameter to 1 in all our experiments.

\FloatBarrier
\section{Terminology}
\label{sec:terminology}
\begin{table*}
    \centering
    \caption{Notation.}
    \small
    \begin{tabular}{|c|p{9cm}|}
        \hline
        term & explanation \\
        \hline
        $\mathbf{s}$ & Input spectrum. \\ 
        $\mathbf{l}$ & Ground truth labels (spectra parameters). \\
        $\hat{\mathbf{s}}$ & Spectrum prediction. \\ 
        $\hat{\mathbf{l}}$ & Label prediction. \\          
        $\mu^k$ & The mean value of the $k$th label. \\   
        $\sigma^k$ & The standard deviation of the $k$th label. \\
        $\underline{\mathbf{l}}$ & Normalized labels according to Eq.~\ref{eq:norm}. \\
        $\mathbf{l}_{\text{scale}}$ & Scaled labels according to Eq.~\ref{eq:norm}. \\
        $\triangle l^k$ & Intervention for $k$th label. \\
        \hline
        $\hat{\mathbf{l}}^{\text{LIF}}$ & Label-informed factors in the latent representation. They are nodes in the latent representation that are supervised with ground truth labels $\mathbf{l}$ during training. They are treated as label prediction $\hat{\mathbf{l}}$ during inference. \\ 
        $\mathbf{u}$ & Unsupervised factors in the latent representation. Associated nodes represent undetermined spectra parameters and statistically relevant features. \\
        $\mathbf{b}$ & Latent representation is a concatenation of $\hat{\mathbf{l}}^{\text{LIF}}$ and $\mathbf{u}$. \\
        $\mathcal{B}$ Latent space.\\
        $M: \mathbf{b} \rightarrow \mathbf{s}$  & Hidden generative process that maps latent representation to the spectrum. \\
        \hline
        $D$ & Number of samples in the HARPS training dataset. \\
        $N$ & Number of pixels in the HARPS or ETC spectrum. \\
        $K$ & Number of supervised and injected labels. \\
        \hline
        $\mathbf{q}^c$ & Classic encoder---part of autoencoder (AE)---projects input spectrum into latent representation $\mathbf{b}$. \\
        $\mathbf{q}^p$ & Probabilistic encoder---part of variational autoencoder (VAE)---that projects input spectrum into distribution over latent representations. \\ 
        $\mathbf{\mu}$ & Mean values of latent representation $\mathbf{b}$. \\
        $\mathbf{\sigma}$ & Standard deviation of latent representation $\mathbf{b}$. \\
        $\mathbf{p}$ & Decoder---used by both VAE and AE---that projects latent representation $\mathbf{b}$ to output spectrum $\hat{\mathbf{s}}$. \\
        \hline
        $L$ & Loss function. \\
        $L_{\text{rec}}$ & Reconstruction loss function. \\
        $L_{\text{lab}}$ & Label loss function. \\
        $D_{\mathbb{KL}}$ & Kullback-Leibler divergence (also known as information gain or relative entropy). \\
        $\lambda_{\text{lab}}$ & Label loss weight. \\
        $\lambda_{\mathbb{KL}}$ & KL loss weight. \\
        \hline
        $\doop(\mathbf{b}, \triangle l^k, k)$ & $\doop$ operator returns modified $\mathbf{b'}$, where $\triangle l^k$ is added to the $k$th element of $\mathbf{b}$.\\
        $\shift(\mathbf{s}, v)$ & Doppler shifts of the spectrum $\mathbf{s}$ by radial velocity $v$.\\
        $\circ$&
        Function composition operator, $g(f(x))=(g\circ f)(x)$.\\
        \hline
        $\mathbb{E}$ & Expected value.\\
        \hline
    \end{tabular}
    \label{tab:notation}
\end{table*}

\begin{table*}
    \centering
    \caption{Terminology}
    \small
    \begin{tabular}{|c|p{9cm}|}
        \hline
        term & explanation \\
        \hline
        spectral parameters, labels & These are parametric descriptions of a
        spectrum, providing quantifiable metrics that characterize its
        properties. In the context of our study, spectral parameters include
        elements such as radial velocity, temperature, metallicity, gravity,
        airmass, and water vapor \citep{stellar_find, telluric_find}.\\
        \hline
        ML parameters & ML parameters refer to the elements of a ML model that we fit to the training data.
        In the context of neural networks, this typically includes the internal weights and biases that are adjusted through training to minimize loss function.\\
        ML hyperparameters & ML hyperparameters are the remaining parameters that complete the ML model description, which are not tuned during training.
        They include choices related to architecture, learning rate, batch size, and any other aspects not covered by the model parameters.
        We search for the optimal ML hyperparameters by comparing ML models with different sets of ML hyperparameters by evaluating the loss function on the validation dataset.
          \\
        \hline
        factor & A factor refers to an individual, independent source of variation within the dataset. Our models aim to identify and isolate these factors for a more interpretable and comprehensible data representation.\\
        latent representation & A compressed representation of the input data.\\
        bottleneck & A narrow layer of AEs that provides latent representations. \\
        latent space & A space consisting of latent representations.\\
        \hline
        label-aware & Models operating under supervised or semi-supervised paradigms.\\
        \hline
        reconstruction & The goal of reconstruction task
        is to reconstruct an input $\mathbf{s}$ from latent representation $\mathbf{b}$. \\
        disentanglement & Disentangled representation separates the underlying factors.  \\
        factor injection & Directly supervising bottleneck with a known factors. \\
        \hline
        out-of-distribution data & Data that have a different distribution than the training dataset. \\
        \hline
        transfer learning & Learning from one task and applying it to
        another task. For example, training a model on the ETC HARPS dataset and
        applying it to a real HARPS dataset. \\
        \hline
        one-shot learning & Transfer learning from a single example per class.
        This is similar to how humans learn. \\
        \hline
        zero-shot learning & Transfer learning without any new training data.
        We train only on real HARPS parameters and apply it to the ETC HARPS dataset or vice versa. \\
        \hline 
        downstream learning & Using the learned representation $\mathbf{b}$ for another task (like source parameters prediction). Downstream learning can provide a metric to evaluate the effectiveness of the data compression. \\
        \hline
    \end{tabular}
    \label{tab:terminology}
\end{table*}

\end{appendix}

\end{document}